\documentclass[a4paper,onecolumn,11pt,unpublished,noarxiv,allowtoday]{quantumarticle}
\usepackage[utf8]{inputenc}
\usepackage[english]{babel}
\usepackage[T1]{fontenc}
\usepackage{amsmath}
\usepackage{hyperref}
\usepackage{mathtools, nccmath}
\usepackage{setspace}
\usepackage{hyperref}
\usepackage{outlines}
\usepackage{fancyvrb}
\usepackage{mathtools}
\usepackage{physics}
\usepackage{comment}
\usepackage{bbold}
\usepackage{amssymb}
\usepackage{cancel}
\usepackage{graphicx}
\usepackage[caption=false]{subfig}
\usepackage[usenames, dvipsnames]{color}
\usepackage{soul}
\usepackage{tensor}
\usepackage{amsthm}
\usepackage[caption=false]{subfig}
\usepackage{algorithm}
\usepackage{algcompatible}

\usepackage{tikz}
\usepackage{lipsum}
\usepackage{tkz-graph}
\usetikzlibrary{hobby,backgrounds,calc,trees}
\usepackage{xcolor}
\usepackage{soul}
\usetikzlibrary{positioning}
\usepackage{ulem}
\usepackage{yquant}
\usepackage{caption}
\usepackage{subfig}
\useyquantlanguage{groups}
\usepackage{makecell}
\usepackage{tikz,esvect,tikz-3dplot}
\usetikzlibrary{3d,calc,intersections}
\usepackage{tikz}
\usepackage{tikz-3dplot}
\usepackage{adjustbox}
\usepackage{zx-calculus}
\usetikzlibrary{zx-calculus}
\usepackage{tikz-cd}
\usepackage[official]{eurosym}

\makeatletter

\usepackage{tensor}
\usepackage{accents}
\usetikzlibrary{angles,quotes}

\definecolor{zx_red}{RGB}{232, 165, 165}
\definecolor{zx_green}{RGB}{216, 248, 216}

\usepackage[numbers,sort&compress]{natbib}
\usepackage{graphicx}
\usepackage{iftex}
\usepackage{braket}
\usepackage{tikzit}
\usepackage{amsfonts,amssymb}
\usepackage{tablefootnote}

\tikzstyle{gate}=[shape=rectangle, text height=1.5ex, text depth=0.25ex, yshift=0.5mm, fill=white, draw=black, minimum height=5mm, yshift=-0.5mm, minimum width=5mm, font={\small}, tikzit category=circuit]
\tikzstyle{big gate}=[shape=rectangle, text height=1.5ex, text depth=0.25ex, yshift=0.5mm, fill=white, draw=black, minimum height=10mm, yshift=-0.5mm, minimum width=5mm, font={\small}, tikzit category=circuit]
\tikzstyle{Z dot}=[inner sep=0mm, minimum size=2mm, shape=circle, draw=black, fill={rgb,255: red,221; green,255; blue,221}, tikzit category=zx]
\tikzstyle{Z phase dot}=[minimum size=5mm, font={\footnotesize\boldmath}, shape=rectangle, rounded corners=2mm, inner sep=0.2mm, outer sep=-2mm, scale=0.8, tikzit shape=rectangle, draw=black, fill={rgb,255: red,221; green,255; blue,221}, tikzit draw=blue, tikzit category=zx]
\tikzstyle{X dot}=[Z dot, shape=circle, draw=black, fill={rgb,255: red,255; green,136; blue,136}, tikzit category=zx]
\tikzstyle{X phase dot}=[Z phase dot, tikzit shape=rectangle, tikzit draw=blue, fill={rgb,255: red,255; green,136; blue,136}, font={\footnotesize\boldmath}, tikzit category=zx]
\tikzstyle{hadamard}=[fill=yellow, draw=black, shape=rectangle, inner sep=0.6mm, minimum height=1.5mm, minimum width=1.5mm, tikzit category=zx]
\tikzstyle{paulibox}=[fill={rgb,255: red,221; green,221; blue,255}, draw=black, shape=rectangle, inner sep=0.6mm, minimum height=5mm, minimum width=5mm, font={\footnotesize}, text height=1.5ex, text depth=0.25ex, tikzit category=zx]
\tikzstyle{vertex}=[inner sep=0mm, minimum size=1mm, shape=circle, draw=black, fill=black, tikzit category=misc]
\tikzstyle{vertex set}=[inner sep=0mm, minimum size=1mm, shape=circle, draw=black, fill=white, font={\footnotesize\boldmath}, tikzit category=misc]
\tikzstyle{small black dot}=[fill=black, draw=black, shape=circle, inner sep=0pt, minimum width=1.2mm, tikzit category=circuit]
\tikzstyle{cnot ctrl}=[fill=black, draw=black, shape=circle, inner sep=0pt, minimum width=1.2mm, tikzit category=circuit]
\tikzstyle{cnot targ}=[fill=white, draw=white, shape=circle, tikzit category=circuit, label={center:$\oplus$}, inner sep=0pt, minimum width=2.1mm, tikzit fill={rgb,255: red,102; green,204; blue,255}, tikzit draw=black]
\tikzstyle{ket}=[fill=white, draw=black, shape=regular polygon, regular polygon sides=3, regular polygon rotate=-30, scale=0.7, inner sep=1pt, tikzit category=circuit, tikzit shape=rectangle, tikzit fill=green]
\tikzstyle{bra}=[fill=white, draw=black, shape=regular polygon, regular polygon sides=3, regular polygon rotate=30, scale=0.7, inner sep=1pt, tikzit category=circuit, tikzit shape=rectangle, tikzit fill=red]
\tikzstyle{scalar}=[shape=rectangle, text height=1.5ex, text depth=0.25ex, yshift=0.5mm, fill=white, draw=black, minimum height=5mm, yshift=-0.5mm, minimum width=5mm, font={\small}]
\tikzstyle{clabel}=[fill=white, draw=none, shape=rectangle, tikzit fill={rgb,255: red,56; green,255; blue,242}, font={\footnotesize}, inner sep=1pt, tikzit category=labels]
\tikzstyle{empty diagram}=[draw={gray!40!white}, dashed, shape=rectangle, minimum width=1cm, minimum height=1cm, tikzit category=misc]
\tikzstyle{amap}=[fill=white, draw=black, shape=NEbox, tikzit category=asymmetric, tikzit fill=yellow, tikzit shape=rectangle]
\tikzstyle{amap conj}=[fill=white, draw=black, shape=NWbox, tikzit category=asymmetric, tikzit fill=green, tikzit shape=rectangle]
\tikzstyle{amap adj}=[fill=white, draw=black, shape=SEbox, tikzit category=asymmetric, tikzit fill=red, tikzit shape=rectangle]
\tikzstyle{amap trans}=[fill=white, draw=black, shape=SWbox, tikzit category=asymmetric, tikzit fill=orange, tikzit shape=rectangle]
\tikzstyle{astate}=[fill=white, draw=black, shape=NEtriangle, tikzit category=asymmetric, tikzit shape=circle, tikzit fill=yellow]
\tikzstyle{astate conj}=[fill=white, draw=black, shape=NWtriangle, tikzit category=asymmetric, tikzit shape=circle, tikzit fill=green]
\tikzstyle{astate adj}=[fill=white, draw=black, shape=SEtriangle, tikzit category=asymmetric, tikzit shape=circle, tikzit fill=red]
\tikzstyle{astate trans}=[fill=white, draw=black, shape=SWtriangle, tikzit category=asymmetric, tikzit shape=circle, tikzit fill=orange]

\tikzstyle{hadamard edge}=[-, dashed, dash pattern=on 2pt off 0.5pt, thick, draw={rgb,255: red,68; green,136; blue,255}]
\tikzstyle{box edge}=[-, dashed, dash pattern=on 2pt off 0.5pt, thick, draw={rgb,255: red,203; green,192; blue,225}]
\tikzstyle{brace edge}=[-, tikzit draw=blue, decorate, decoration={brace,amplitude=1mm,raise=-1mm}]
\tikzstyle{diredge}=[->]
\tikzstyle{double edge}=[-, double, shorten <=-1mm, shorten >=-1mm, double distance=2pt]
\tikzstyle{gray edge}=[-, {gray!60!white}]
\tikzstyle{pointer edge}=[->, very thick, gray]
\tikzstyle{boldedge}=[-, line width=1.6pt, shorten <=-0.17mm, shorten >=-0.17mm]
\tikzstyle{bidir edge}=[<->, very thick, draw={rgb,255: red,191; green,191; blue,191}]
\tikzstyle{separator edge}=[-, dashed, dash pattern=on 2pt off 0.5pt, thick, draw={rgb,255: red,153; green,153; blue,153}]

\begin{document}

\title{Cutting stabiliser decompositions of magic state cultivation with ZX-calculus}

\date{\today}

\author{Kwok Ho Wan}
\orcid{0000-0002-1762-1001}
\affiliation{Blackett Laboratory, Imperial College London, South Kensington, London SW7 2AZ, UK}
\affiliation{Mathematical Institute, University of Oxford, Andrew Wiles Building, Woodstock Road, Oxford OX2 6GG, UK}

\author{Zhenghao Zhong}
\orcid{0000-0001-5159-1013}
\affiliation{Mathematical Institute, University of Oxford, Andrew Wiles Building, Woodstock Road, Oxford OX2 6GG, UK}
\affiliation{Blackett Laboratory, Imperial College London, South Kensington, London SW7 2AZ, UK}

\begin{abstract}
We apply the cutting stabiliser decomposition techniques [arXiv:2403.10964] to the quantum states generated from magic state cultivation [arXiv:2409.17595], post-selected upon all $+1$ measured values for simplicity. The resultant states to the $d=3$ and $d=5$ variant magic state cultivation circuits can be expressed as a sum of $4$ and $8$ Clifford ZX-diagrams respectively. Modifications to existing ZX-calculus stabiliser decomposition methods may enable better simulation of non-Clifford circuits containing a moderate number of $T$ gates in the context of quantum error correction.
\end{abstract}

\maketitle

\section{Introduction}
The presence of non-Clifford $T = \begin{pmatrix}
1 & 0 \\
0 & e^{i\pi/4}
\end{pmatrix}$ gates in a quantum circuit poses challenges for \textit{strong} classical simulations \cite{gottesman1998heisenbergrepresentationquantumcomputers,Aaronson_2004}. For an arbitrary $n$-qubit quantum circuit acting on $\ket{0}^{\otimes n}$ with $t$ number of $T$ gates, the classical simulation space-time complexity is $\mathcal{O}(2^{\alpha t}\text{poly}(n))$, for some constant $\alpha>0$ \cite{Aaronson_2004,PhysRevLett.116.250501}. Due to conjectures in complexity theory, efficient exact classical simulation of quantum circuits is highly unlikely \cite{Bremner_2010}. On the other hand, stabiliser/Clifford states evolved under Clifford unitaries and measurements are considered easy and polynomially tractable to simulate via the Gottesman-Knill theorem \cite{Aaronson_2004,gottesman1998heisenbergrepresentationquantumcomputers}. 

There are various ways of tackling classical simulation of quantum circuits with small to moderate number of $T$ gates. One approach is to decompose the corresponding non-Clifford state as a superposition of Clifford states. A general $n$-qubit quantum state ($\ket{\Psi}$) can be decomposed as a sum of $\chi$ number of Clifford states ($\ket{\phi_j}$):
\begin{equation}
    \ket{\Psi} = \sum_{j=1}^{\chi} a_j \ket{\phi_j} \ ,
\end{equation}

\noindent for some complex coefficients $a_j$. In general, the number of terms in the sum follows the scaling of: $\chi \sim \mathcal{O}(2^{\alpha t})$, where $\alpha>0$ and $t$ is the number of $T$ gates used in a quantum circuit $U$ needed to construct said state ($\ket{\Psi} = U \ket{0}^{\otimes n}$). The objective of stabiliser decomposition is to find a low $\alpha$ representation of the state $\ket{\Psi}$. At moderately low $T$-count\footnote{Number of $\ket{T}$-states or $T$-gates.} ($\sim 30$ to $50$), circuits with special structures may still admit stabiliser decompositions with relatively few terms, compared to the worst case scaling of $2^t$. This may indicate potentially favourable requirements for simulation. In addition, a single magic $\ket{T} \propto \ket{0} + e^{i\pi/4}\ket{1}$ states can be consumed to perform a $T$ gate where the $T$-count here means the number of $\ket{T}$ states or $T$ gates involved interchangeably.

The aim of this manuscript is to present the state-of-the-art stabiliser decomposition techniques \cite{Sutcliffe_2024} applied to the zero-level magic state distillation circuits \cite{gidney2024magicstatecultivationgrowing}, coined as magic state cultivation (MSC). Logical magic states are resources in fault tolerant quantum computation. MSC promises high fidelity ($10^{-9}$ logical error rate at circuit level noise of $0.001$) with sufficiently low $10^3$ expected qubit-rounds spacetime volume on a planar local qubit architecture. The logical magic states stemming from MSC could potentially make multi-level magic state distillation redundant \cite{Bravyi_2005, Litinski_2019}. However, the high number of $T$ gates involved in MSC makes it difficult for classical simulations. 

We found that states generated from the $d=3$ and $d=5$ colour code variants of the MSC circuit can be represented by sums of $4$ and $8$ Clifford ZX-diagrams respectively via the cutting decomposition.

\textbf{Disclaimer:} We do not claim that we can simulate the MSC circuit from end-to-end with arbitrary Pauli errors and different syndrome detection patterns. We are aware that the resultant state to both the $d=3$ and $d=5$ MSC circuits are just encoded magic states: $\ket{\bar{T}} \propto \ket{\bar{0}} + e^{i\pi/4}\ket{\bar{1}}$, a superposition of \textbf{$2$ Clifford terms}, beating the number of terms obtained by the cutting decomposition. Nonetheless, it felt like an interesting exercise to perform the cutting decomposition from beginning to end. That said, we hope that more stabiliser decomposition techniques can be modified to allow for better classical simulations and verifications of MSC. In other words, this is an early idea and preliminary work\footnote{In other words, we have moved onto new jobs.}.

We encourage any potential readers to load the $\mathtt{.tikz}$ files from this article's arXiv tex source into $\mathtt{pyzx}$/$\mathtt{zxlive}$ \cite{kissinger2020Pyzx} and interact with the diagrams.

\section{ZX-diagram based stabiliser decompositions}
ZX-calculus will be the preferred way of representing various good stabiliser decompositions in this manuscript. For the sake of keeping this manuscript shorter, we assume the reader has a basic understanding of ZX-calculus and stabiliser calculations \cite{Aaronson_2004}. For a review on ZX(H)-calculus applied in the context of quantum software, please refer to \cite{KissingerWetering2024Book}. We also want to note that most of the ZX TikZ diagram in this section are from \cite{Kiss2022,Sutcliffe_2024}.

Before decomposing  $t$ tensored copies of magic states: $\ket{T}^{\otimes t} = \Big(\scalebox{1}{
\begin{tikzpicture}
    \begin{pgfonlayer}{nodelayer}
        \node [style=Z phase dot] (0) at (-2.00, -0.00) {$\frac{\pi}{4}$};
        \node [style=none] (1) at (-1.00, -0.00) {};
    \end{pgfonlayer}
    \begin{pgfonlayer}{edgelayer}
        \draw (0) to (1);
    \end{pgfonlayer}
\end{tikzpicture}
}\Big)^{\otimes t}$ into fewer than $2^t$ terms, we first look at the decomposition of  $\ket{T}^{\otimes 2}$ into $2$ rather than $2^2=4$ terms \cite{Qassim_2021}:
\begin{equation}
    \label{eq:T2}
    \ket{T}^{\otimes 2} \propto \overbrace{(\ket{00}+i\ket{11})}^{\text{Clifford}} + e^{i\pi/4}\underbrace{(\ket{01}+\ket{10})}_{\text{Clifford}} \ .
\end{equation}

\noindent The equivalent ZX-diagram representation of \eqref{eq:T2} is:
\begin{equation}
    \label{eq:2-T-decomposition}
    \scalebox{1}{\begin{tikzpicture}
	\begin{pgfonlayer}{nodelayer}
		\node [style=Z phase dot] (0) at (-1.5, 0.488) {$\frac\pi4$};
		\node [style=none] (1) at (-0.475, 0.487) {};
		\node [style=Z phase dot] (2) at (-1.475, -0.512) {$\frac\pi4$};
		\node [style=none] (3) at (-0.45, -0.513) {};
		\node [style=none] (4) at (0.5, 0) {$=$};
		\node [style=Z phase dot] (5) at (1.75, -0.012) {$\frac\pi2$};
		\node [style=none] (6) at (2.775, 0.487) {};
		\node [style=none] (8) at (2.8, -0.513) {};
		\node [style=none] (9) at (4, 0) {$+\ e^{i\frac\pi4}$};
		\node [style=Z phase dot] (10) at (5.5, 0) {$\pi$};
		\node [style=Z dot] (11) at (6.75, 0.5) {};
		\node [style=Z dot] (12) at (6.75, -0.5) {};
		\node [style=none] (13) at (7.5, 0.5) {};
		\node [style=none] (14) at (7.5, -0.5) {};
	\end{pgfonlayer}
	\begin{pgfonlayer}{edgelayer}
		\draw (1.center) to (0);
		\draw (3.center) to (2);
		\draw [bend right] (6.center) to (5);
		\draw [bend right] (5) to (8.center);
		\draw (13.center) to (11);
		\draw (14.center) to (12);
		\draw [style=hadamard edge] (10) to (12);
		\draw [style=hadamard edge] (10) to (11);
	\end{pgfonlayer}
\end{tikzpicture}} \ ,
\end{equation}

\noindent where blue dotted lines mean an hadamard gate on the edge ($\scalebox{1}{\begin{tikzpicture}
    \begin{pgfonlayer}{nodelayer}
        \node [style=none] (0) at (-1.00, -0.00) {};
        \node [style=hadamard] (1) at (-0.00, -0.00) {};
        \node [style=none] (2) at (1.00, -0.00) {};

        \node [] (4) at (2.00, -0.00) {$=$};
        \node [style=none] (5) at (3.00, -0.00) {};
        \node [style=none] (6) at (5.00, -0.00) {};
    \end{pgfonlayer}
    \begin{pgfonlayer}{edgelayer}
        \draw (0) to (1);
        \draw (2) to (1);
        \draw [style=hadamard edge] (5) to (6);
    \end{pgfonlayer}
\end{tikzpicture}}$). 

The following is a quick non-exhaustive review on good stabiliser decompositions at low $t$. We first go over each example as sums of Clifford ZX-diagrams. Then, we will focus on using the cutting decomposition for MSC circuits in the next sections. Here are three different ZX-calculus based stabiliser decompositions:
\begin{enumerate}
    \item Decomposition from Bravyi Smith and Smolin (BSS) \cite{BSS2016}, 
    \item Magic cat states decomposition from \cite{Qassim_2021}, and subsequently generalised in the language of ZX-diagrams \cite{Kiss2022,Codsi2022Masters},
    \item Cutting decomposition \cite{Codsi2022Masters,Sutcliffe2025thesis,Sutcliffe_2024}.
\end{enumerate}

\subsection{BSS decomposition}
The BSS decomposition can be written as ZX-diagrams:
\begin{equation}
    \label{eq:bss_decomp}
    \scalebox{0.7}{\input{bss_decomp.tikz}} \ .
\end{equation}

\noindent where $6$ magic $T$ states can be represented as a sum of $7$ Clifford ZX-diagram. A naive multiplicative decomposition of $\ket{T}^{\otimes 6}$ will yield $2^6 = 64$ terms.

\subsection{Magic cat state decomposition}
Magic cat states parametrised by positive integer $m$ are defined by:
\begin{equation}
    \ket{\text{cat}_m} = \frac{1}{\sqrt{2}} (I^{\otimes m}+Z^{\otimes m})\ket{T}^{\otimes m} \ ,
\end{equation}

\noindent where $I$ is the identity operator on the support of a qubit and $Z$ is the Pauli-Z operator. Various $\ket{\text{cat}_m}$ stabiliser decompositions exist, please refer to \cite{Qassim_2021,Codsi2022Masters,Kiss2022} for further details. A $\ket{\text{cat}_m}$ can be written as the following ZX-diagram
\begin{equation}
    \label{eq:cat-n}
    \scalebox{1}{\begin{tikzpicture}[scale=2]
	\begin{pgfonlayer}{nodelayer}
		\node [style=Z phase dot] (0) at (-0.113, 0.65) {$\frac\pi4$};
		\node [style=none] (1) at (0.238, 0.65) {};
		\node [style=Z phase dot] (3) at (-0.113, 0) {$\frac\pi4$};
		\node [style=none] (4) at (0.238, 0) {};
		\node [style=Z phase dot] (6) at (-0.113, -1.15) {$\frac\pi4$};
		\node [style=none] (7) at (0.238, -1.15) {};
		\node [style=Z dot] (9) at (-0.75, -0.25) {};
		\node [style=none] (10) at (-0.125, -0.475) {\vdots};
	\end{pgfonlayer}
	\begin{pgfonlayer}{edgelayer}
		\draw (1.center) to (0);
		\draw (4.center) to (3);
		\draw (7.center) to (6);
		\draw [style=hadamard edge, in=-165, out=90] (9) to (0);
		\draw [style=hadamard edge, in=-180, out=45] (9) to (3);
		\draw [style=hadamard edge, in=165, out=-90] (9) to (6);
	\end{pgfonlayer}
\end{tikzpicture}} \ ,
\end{equation}

\noindent with $m$ legs. We shall illustrate the $\ket{\text{cat}_6}$ decomposition here:
\begin{equation}
    \label{eq:cat-6-decomp}
    \scalebox{0.7}{\begin{tikzpicture}[xscale=1.7, yscale=2]
	\begin{pgfonlayer}{nodelayer}
		\node [style=Z phase dot] (0) at (-1.363, -1.425) {$\frac\pi4$};
		\node [style=none] (1) at (-0.762, -1.425) {};
		\node [style=Z phase dot] (3) at (-1.363, -0.825) {$\frac\pi4$};
		\node [style=none] (4) at (-0.762, -0.825) {};
		\node [style=Z dot] (9) at (-2, 0.05) {};
		\node [style=Z phase dot] (10) at (-1.363, -0.225) {$\frac\pi4$};
		\node [style=none] (11) at (-0.762, -0.225) {};
		\node [style=Z phase dot] (13) at (-1.363, 0.35) {$\frac\pi4$};
		\node [style=none] (14) at (-0.762, 0.35) {};
		\node [style=Z phase dot] (16) at (-1.363, 0.95) {$\frac\pi4$};
		\node [style=none] (17) at (-0.762, 0.95) {};
		\node [style=Z phase dot] (19) at (-1.363, 1.55) {$\frac\pi4$};
		\node [style=none] (20) at (-0.762, 1.55) {};
		\node [style=none] (22) at (0.25, 0) {$=$};
		\node [style=none] (23) at (2.512, -1.45) {};
		\node [style=none] (26) at (2.512, -0.85) {};
		\node [style=Z phase dot] (29) at (1.625, 0.025) {$~-\frac\pi2~$};
		\node [style=none] (30) at (2.512, -0.25) {};
		\node [style=none] (33) at (2.512, 0.325) {};
		\node [style=none] (36) at (2.512, 0.925) {};
		\node [style=none] (39) at (2.512, 1.525) {};
		\node [style=none] (41) at (3.75, 0) {$+~\frac{i e^{i\pi/4}}{\sqrt2}$};
		\node [style=Z dot] (42) at (5.512, -1.425) {};
		\node [style=Z dot] (43) at (5.512, -0.825) {};
		\node [style=Z dot] (44) at (4.875, 0.05) {};
		\node [style=Z dot] (45) at (5.512, -0.225) {};
		\node [style=Z dot] (46) at (5.512, 0.35) {};
		\node [style=Z dot] (47) at (5.512, 0.95) {};
		\node [style=Z dot] (48) at (5.512, 1.55) {};
		\node [style=none] (49) at (6.75, 0) {$-\ \frac{e^{i\pi/4}}{\sqrt{2}}$};
		\node [style=none] (68) at (5.887, -1.425) {};
		\node [style=none] (69) at (5.887, -0.825) {};
		\node [style=none] (70) at (5.887, -0.225) {};
		\node [style=none] (71) at (5.887, 0.35) {};
		\node [style=none] (72) at (5.887, 0.95) {};
		\node [style=none] (73) at (5.887, 1.55) {};
		\node [style=none] (74) at (0.75, 0) {$\frac12$};
		\node [style=Z phase dot] (75) at (8.512, -1.425) {$\frac\pi2$};
		\node [style=Z phase dot] (76) at (8.512, -0.825) {$\frac\pi2$};
		\node [style=Z dot] (77) at (7.625, 0.05) {};
		\node [style=Z phase dot] (78) at (8.512, -0.225) {$\frac\pi2$};
		\node [style=Z phase dot] (79) at (8.512, 0.35) {$\frac\pi2$};
		\node [style=Z phase dot] (80) at (8.512, 0.95) {$\frac\pi2$};
		\node [style=Z phase dot] (81) at (8.512, 1.55) {$\frac\pi2$};
		\node [style=none] (82) at (9.012, -1.425) {};
		\node [style=none] (83) at (9.012, -0.825) {};
		\node [style=none] (84) at (9.012, -0.225) {};
		\node [style=none] (85) at (9.012, 0.35) {};
		\node [style=none] (86) at (9.012, 0.95) {};
		\node [style=none] (87) at (9.012, 1.55) {};
	\end{pgfonlayer}
	\begin{pgfonlayer}{edgelayer}
		\draw (1.center) to (0);
		\draw (4.center) to (3);
		\draw [style=hadamard edge, in=165, out=-90] (9) to (0);
		\draw [style=hadamard edge, in=165, out=-75] (9) to (3);
		\draw (11.center) to (10);
		\draw (14.center) to (13);
		\draw (17.center) to (16);
		\draw (20.center) to (19);
		\draw [style=hadamard edge, in=180, out=-45] (9) to (10);
		\draw [style=hadamard edge, in=45, out=180] (13) to (9);
		\draw [style=hadamard edge, in=-165, out=75] (9) to (16);
		\draw [style=hadamard edge, in=-165, out=90] (9) to (19);
		\draw [in=165, out=-90] (29) to (23.center);
		\draw [in=165, out=-75] (29) to (26.center);
		\draw [in=180, out=-45] (29) to (30.center);
		\draw [in=45, out=180] (33.center) to (29);
		\draw [in=-165, out=75] (29) to (36.center);
		\draw [in=-165, out=90] (29) to (39.center);
		\draw [style=hadamard edge, in=165, out=-90] (44) to (42);
		\draw [style=hadamard edge, in=165, out=-75] (44) to (43);
		\draw [style=hadamard edge, in=180, out=-45] (44) to (45);
		\draw [style=hadamard edge, in=45, out=180] (46) to (44);
		\draw [style=hadamard edge, in=-165, out=75] (44) to (47);
		\draw [style=hadamard edge, in=-165, out=90] (44) to (48);
		\draw (42) to (68.center);
		\draw (43) to (69.center);
		\draw (45) to (70.center);
		\draw (46) to (71.center);
		\draw (47) to (72.center);
		\draw (48) to (73.center);
		\draw [style=hadamard edge, in=165, out=-90] (77) to (75);
		\draw [style=hadamard edge, in=165, out=-75] (77) to (76);
		\draw [style=hadamard edge, in=180, out=-45] (77) to (78);
		\draw [style=hadamard edge, in=45, out=180] (79) to (77);
		\draw [style=hadamard edge, in=-165, out=75] (77) to (80);
		\draw [style=hadamard edge, in=-165, out=90] (77) to (81);
		\draw (75) to (82.center);
		\draw (76) to (83.center);
		\draw (78) to (84.center);
		\draw (79) to (85.center);
		\draw (80) to (86.center);
		\draw (81) to (87.center);
	\end{pgfonlayer}
\end{tikzpicture}} \ .
\end{equation}

We want to note that a $T$ measurement can be performed on any outward leg of a $\ket{\text{cat}_m}$ state in general to produce a $\ket{T}^{\otimes (m-1)} = (\scalebox{1}{
\begin{tikzpicture}
    \begin{pgfonlayer}{nodelayer}
        \node [style=Z phase dot] (0) at (-2.00, -0.00) {$\frac{\pi}{4}$};
        \node [style=none] (1) at (-1.00, -0.00) {};
    \end{pgfonlayer}
    \begin{pgfonlayer}{edgelayer}
        \draw (0) to (1);
    \end{pgfonlayer}
\end{tikzpicture}
})^{\otimes (m-1)}$ state. This can be seen in \eqref{eq:magic-from-cat} for the $\ket{\text{cat}_6}$ state, resulting in a $\ket{T}^{\otimes 5}$ decomposition consisting of $3$ (almost-Clifford\footnote{We can see that each term in the stabiliser decomposition of \eqref{eq:magic-from-cat} on the RHS has a single $\pm\pi/4$ spider, hence the total number of Clifford ZX-diagram is $6$ using \eqref{eq:single_magic}.}) terms instead of $2^5=32$ terms.
\begin{equation}
    \label{eq:magic-from-cat}
    \scalebox{0.7}{\begin{tikzpicture}[scale=2]
	\begin{pgfonlayer}{nodelayer}
		\node [style=Z phase dot] (0) at (1.812, 1.55) {$\frac\pi4$};
		\node [style=Z phase dot] (1) at (2.613, 1.55) {$~-\frac\pi4~$};
		\node [style=Z phase dot] (3) at (1.812, 0.925) {$\frac\pi4$};
		\node [style=none] (4) at (2.163, 0.925) {};
		\node [style=Z dot] (6) at (1.175, 0) {};
		\node [style=Z phase dot] (7) at (1.812, 0.3) {$\frac\pi4$};
		\node [style=none] (8) at (2.163, 0.3) {};
		\node [style=Z phase dot] (10) at (1.812, -0.325) {$\frac\pi4$};
		\node [style=none] (11) at (2.163, -0.325) {};
		\node [style=Z phase dot] (13) at (1.812, -0.95) {$\frac\pi4$};
		\node [style=none] (14) at (2.163, -0.95) {};
		\node [style=Z phase dot] (16) at (1.812, -1.575) {$\frac\pi4$};
		\node [style=none] (17) at (2.163, -1.575) {};
		\node [style=none] (98) at (0.05, 0) {$=$};
		\node [style=Z phase dot] (105) at (-0.863, -1.625) {$\frac\pi4$};
		\node [style=none] (106) at (-0.337, -1.625) {};
		\node [style=Z phase dot] (107) at (-0.863, -0.875) {$\frac\pi4$};
		\node [style=none] (108) at (-0.337, -0.875) {};
		\node [style=Z phase dot] (109) at (-0.863, -0.125) {$\frac\pi4$};
		\node [style=none] (110) at (-0.337, -0.125) {};
		\node [style=Z phase dot] (111) at (-0.863, 0.625) {$\frac\pi4$};
		\node [style=none] (112) at (-0.337, 0.625) {};
		\node [style=Z phase dot] (113) at (-0.863, 1.375) {$\frac\pi4$};
		\node [style=none] (114) at (-0.337, 1.375) {};
		\node [style=none] (120) at (0.6, 0) {$4$};
		\node [style=none] (134) at (5.062, -1.525) {};
		\node [style=none] (135) at (5.062, -0.9) {};
		\node [style=Z phase dot] (136) at (4.175, 0.025) {$~-\frac\pi2~$};
		\node [style=none] (137) at (5.062, -0.275) {};
		\node [style=none] (138) at (5.062, 0.35) {};
		\node [style=none] (139) at (5.062, 0.975) {};
		\node [style=Z phase dot] (140) at (5.062, 1.6) {$\ -\frac\pi4~$};
		\node [style=none] (141) at (6.3, 0) {$+~2\sqrt{2}i e^{i\pi/4}$};
		\node [style=Z dot] (142) at (8.312, -1.5) {};
		\node [style=Z dot] (143) at (8.312, -0.875) {};
		\node [style=Z dot] (144) at (7.675, 0.05) {};
		\node [style=Z dot] (145) at (8.312, -0.25) {};
		\node [style=Z dot] (146) at (8.312, 0.375) {};
		\node [style=Z dot] (147) at (8.312, 1) {};
		\node [style=Z phase dot] (148) at (8.312, 1.625) {$\ -\frac\pi4~$};
		\node [style=none] (149) at (9.975, 0) {$-\ 2\sqrt{2} e^{i\pi/4}$};
		\node [style=none] (150) at (8.687, -1.5) {};
		\node [style=none] (151) at (8.687, -0.875) {};
		\node [style=none] (152) at (8.687, -0.25) {};
		\node [style=none] (153) at (8.687, 0.375) {};
		\node [style=none] (154) at (8.687, 1) {};
		\node [style=none] (155) at (3.55, 0) {$2$};
		\node [style=Z phase dot] (156) at (11.987, -1.5) {$\frac\pi2$};
		\node [style=Z phase dot] (157) at (11.987, -0.875) {$\frac\pi2$};
		\node [style=Z dot] (158) at (11.1, 0.05) {};
		\node [style=Z phase dot] (159) at (11.987, -0.25) {$\frac\pi2$};
		\node [style=Z phase dot] (160) at (11.987, 0.375) {$\frac\pi2$};
		\node [style=Z phase dot] (161) at (11.987, 1) {$\frac\pi2$};
		\node [style=Z phase dot] (162) at (11.987, 1.625) {$\frac\pi4$};
		\node [style=none] (163) at (12.487, -1.5) {};
		\node [style=none] (164) at (12.487, -0.875) {};
		\node [style=none] (165) at (12.487, -0.25) {};
		\node [style=none] (166) at (12.487, 0.375) {};
		\node [style=none] (167) at (12.487, 1) {};
		\node [style=none] (168) at (2.975, 0) {=};
	\end{pgfonlayer}
	\begin{pgfonlayer}{edgelayer}
		\draw (1) to (0);
		\draw (4.center) to (3);
		\draw [style=hadamard edge, in=-165, out=90] (6) to (0);
		\draw [style=hadamard edge, in=-165, out=75] (6) to (3);
		\draw (8.center) to (7);
		\draw (11.center) to (10);
		\draw (14.center) to (13);
		\draw (17.center) to (16);
		\draw [style=hadamard edge, in=-180, out=45] (6) to (7);
		\draw [style=hadamard edge, in=-45, out=-180] (10) to (6);
		\draw [style=hadamard edge, in=165, out=-75] (6) to (13);
		\draw [style=hadamard edge, in=165, out=-90] (6) to (16);
		\draw (106.center) to (105);
		\draw (108.center) to (107);
		\draw (110.center) to (109);
		\draw (112.center) to (111);
		\draw (114.center) to (113);
		\draw [in=165, out=-90] (136) to (134.center);
		\draw [in=165, out=-75] (136) to (135.center);
		\draw [in=180, out=-45] (136) to (137.center);
		\draw [in=45, out=180] (138.center) to (136);
		\draw [in=-165, out=75] (136) to (139.center);
		\draw [in=-165, out=90] (136) to (140);
		\draw [style=hadamard edge, in=165, out=-90] (144) to (142);
		\draw [style=hadamard edge, in=165, out=-75] (144) to (143);
		\draw [style=hadamard edge, in=180, out=-45] (144) to (145);
		\draw [style=hadamard edge, in=45, out=180] (146) to (144);
		\draw [style=hadamard edge, in=-165, out=75] (144) to (147);
		\draw [style=hadamard edge, in=-165, out=90] (144) to (148);
		\draw (142) to (150.center);
		\draw (143) to (151.center);
		\draw (145) to (152.center);
		\draw (146) to (153.center);
		\draw (147) to (154.center);
		\draw [style=hadamard edge, in=165, out=-90] (158) to (156);
		\draw [style=hadamard edge, in=165, out=-75] (158) to (157);
		\draw [style=hadamard edge, in=180, out=-45] (158) to (159);
		\draw [style=hadamard edge, in=45, out=180] (160) to (158);
		\draw [style=hadamard edge, in=-165, out=75] (158) to (161);
		\draw [style=hadamard edge, in=-165, out=90] (158) to (162);
		\draw (156) to (163.center);
		\draw (157) to (164.center);
		\draw (159) to (165.center);
		\draw (160) to (166.center);
		\draw (161) to (167.center);
	\end{pgfonlayer}
\end{tikzpicture}} \ .
\end{equation}

These $\ket{T}$ states can be consumed to perform $T$ gates in a bigger ZX-diagram. See \cite{Codsi2022Masters} for a table of bigger magic cat state stabiliser decompositions.

\subsection{Cutting}
Stabiliser decomposition by graph cutting is a relatively new technique \cite{Codsi2022Masters,Sutcliffe2025thesis,Sutcliffe_2024}, revolving around this identity:
\begin{equation}
    \label{eq:spider_decomp_abs}
    \scalebox{1}{\begin{tikzpicture}
	\begin{pgfonlayer}{nodelayer}
		\node [style=none] (0) at (-1.25, 1) {};
		\node [style=none] (1) at (-1.25, -1) {};
		\node [style=Z phase dot] (4) at (0, 0) {$\beta$};
		\node [style=none] (6) at (1.25, 1) {};
		\node [style=none] (7) at (1.25, -1) {};
		\node [style=none] (8) at (1.25, 0.25) {$\vdots$};
		\node [style=none] (9) at (-1.25, 0.25) {$\vdots$};
		\node [style=none] (10) at (3, 0) {$\approx$};
		\node [style=none] (11) at (4.75, 1) {};
		\node [style=none] (12) at (4.75, -1) {};
		\node [style=X dot] (13) at (5.75, 0.25) {};
		\node [style=none] (14) at (7.25, 1) {};
		\node [style=none] (15) at (7.25, -1) {};
		\node [style=none] (16) at (7.25, 0.25) {$\vdots$};
		\node [style=none] (17) at (4.75, 0.25) {$\vdots$};
		\node [style=X dot] (18) at (6.25, 0.25) {};
		\node [style=X dot] (19) at (6.25, -0.25) {};
		\node [style=X dot] (20) at (5.75, -0.25) {};
		\node [style=none] (21) at (8.5, 0) {$+$};
		\node [style=none] (22) at (11.5, 1) {};
		\node [style=none] (23) at (11.5, -1) {};
		\node [style=X phase dot] (24) at (12.25, 0.5) {$\pi$};
		\node [style=none] (25) at (14, 1) {};
		\node [style=none] (26) at (14, -1) {};
		\node [style=none] (27) at (14, 0.25) {$\vdots$};
		\node [style=none] (28) at (11.5, 0.25) {$\vdots$};
		\node [style=X phase dot] (29) at (13.25, 0.5) {$\pi$};
		\node [style=X phase dot] (30) at (13.25, -0.5) {$\pi$};
		\node [style=X phase dot] (31) at (12.25, -0.5) {$\pi$};
		\node [style=none] (32) at (10, 0) {$e^{i\beta}$};
	\end{pgfonlayer}
	\begin{pgfonlayer}{edgelayer}
		\draw [bend left] (0.center) to (4);
		\draw [bend left] (4) to (6.center);
		\draw [bend right] (4) to (7.center);
		\draw [bend left] (4) to (1.center);
		\draw [bend left] (11.center) to (13);
		\draw [bend left] (20) to (12.center);
		\draw [bend right] (19) to (15.center);
		\draw [bend left] (18) to (14.center);
		\draw [bend left] (22.center) to (24);
		\draw [bend left] (31) to (23.center);
		\draw [bend right] (30) to (26.center);
		\draw [bend left] (29) to (25.center);
	\end{pgfonlayer}
\end{tikzpicture}} \ ,
\end{equation}
where the Z-spider with phase $\beta$ here is chosen to be cut, resulting in a sum of two ZX-diagram. By choosing spiders to be cut sequentially, carefully in a larger ZX-diagram, each ZX-diagram produced in the stabiliser decomposition sum can be simplified by further ZX-rewrites, to minimise the total number of $\scalebox{1}{
\begin{tikzpicture}
    \begin{pgfonlayer}{nodelayer}
        \node [style=Z phase dot] (0) at (-2.00, -0.00) {$\frac{\pi}{4}$};
        \node [style=none] (1) at (-1.00, -0.00) {};
    \end{pgfonlayer}
    \begin{pgfonlayer}{edgelayer}
        \draw (0) to (1);
    \end{pgfonlayer}
\end{tikzpicture}
}$. For example, see \cite[Section 3.2]{Sutcliffe_2024} reproduced below:
\begin{equation}
    \label{eq:picom_grouped}
    \scalebox{0.7}{\begin{tikzpicture}
	\begin{pgfonlayer}{nodelayer}
		\node [style=none] (0) at (-8, 0) {};
		\node [style=none] (1) at (-8, -1.25) {};
		\node [style=Z phase dot] (2) at (-7, -1.25) {$\frac{\pi}{4}$};
		\node [style=X dot] (3) at (-6, -1.25) {};
		\node [style=Z dot] (4) at (-6, 0) {};
		\node [style=Z phase dot] (5) at (-5, -1.25) {$\frac{\pi}{4}$};
		\node [style=none] (7) at (-1.75, -1.25) {};
		\node [style=none] (62) at (-8, -2.5) {};
		\node [style=Z phase dot] (63) at (-4.75, -2.5) {$\frac{\pi}{4}$};
		\node [style=X dot] (64) at (-3.75, -2.5) {};
		\node [style=Z dot] (65) at (-3.75, 0) {};
		\node [style=Z phase dot] (66) at (-2.75, -2.5) {$\frac{\pi}{4}$};
		\node [style=none] (67) at (-1.75, 0) {};
		\node [style=none] (68) at (-1.75, -2.5) {};
		\node [style=none] (69) at (0.25, 0) {};
		\node [style=none] (70) at (0.25, -1.25) {};
		\node [style=Z phase dot] (71) at (1.25, -1.25) {$\frac{\pi}{4}$};
		\node [style=X dot] (72) at (2.25, -1.25) {};
		\node [style=Z dot] (73) at (3.25, 0) {};
		\node [style=Z phase dot] (74) at (3.25, -1.25) {$\frac{\pi}{4}$};
		\node [style=none] (75) at (6.5, -1.25) {};
		\node [style=none] (76) at (0.25, -2.5) {};
		\node [style=Z phase dot] (77) at (3.5, -2.5) {$\frac{\pi}{4}$};
		\node [style=X dot] (78) at (4.5, -2.5) {};
		\node [style=Z phase dot] (80) at (5.5, -2.5) {$\frac{\pi}{4}$};
		\node [style=none] (81) at (6.5, 0) {};
		\node [style=none] (82) at (6.5, -2.5) {};
		\node [style=none] (83) at (-0.75, -1.25) {$=$};
		\node [style=none] (103) at (8.5, 0) {};
		\node [style=none] (104) at (8.5, -1.25) {};
		\node [style=Z phase dot] (105) at (9.5, -1.25) {$\frac{\pi}{4}$};
		\node [style=X dot] (106) at (10.5, -1.25) {};
		\node [style=Z phase dot] (108) at (11.5, -1.25) {$\frac{\pi}{4}$};
		\node [style=none] (109) at (14.75, -1.25) {};
		\node [style=none] (110) at (8.5, -2.5) {};
		\node [style=Z phase dot] (111) at (11.75, -2.5) {$\frac{\pi}{4}$};
		\node [style=X dot] (112) at (12.75, -2.5) {};
		\node [style=Z phase dot] (113) at (13.75, -2.5) {$\frac{\pi}{4}$};
		\node [style=none] (114) at (14.75, 0) {};
		\node [style=none] (115) at (14.75, -2.5) {};
		\node [style=X dot] (117) at (9.5, 0) {};
		\node [style=X dot] (118) at (13.75, 0) {};
		\node [style=none] (119) at (7.5, -1.25) {$\approx$};
		\node [style=none] (120) at (16.75, 0) {};
		\node [style=none] (121) at (16.75, -1.25) {};
		\node [style=Z phase dot] (122) at (17.75, -1.25) {$\frac{\pi}{4}$};
		\node [style=X dot] (123) at (18.75, -1.25) {};
		\node [style=Z phase dot] (124) at (19.75, -1.25) {$\frac{\pi}{4}$};
		\node [style=none] (125) at (23, -1.25) {};
		\node [style=none] (126) at (16.75, -2.5) {};
		\node [style=Z phase dot] (127) at (20, -2.5) {$\frac{\pi}{4}$};
		\node [style=X dot] (128) at (21, -2.5) {};
		\node [style=Z phase dot] (129) at (22, -2.5) {$\frac{\pi}{4}$};
		\node [style=none] (130) at (23, 0) {};
		\node [style=none] (131) at (23, -2.5) {};
		\node [style=X phase dot] (132) at (17.75, 0) {$\pi$};
		\node [style=X phase dot] (133) at (22, 0) {$\pi$};
		\node [style=none] (134) at (15.75, -1.25) {$+$};
		\node [style=X dot] (135) at (10.5, -0.5) {};
		\node [style=X dot] (136) at (12.75, -1.75) {};
		\node [style=X phase dot] (137) at (18.75, -0.5) {$\pi$};
		\node [style=X phase dot] (138) at (21, -1.75) {$\pi$};
		\node [style=none] (139) at (6.75, -4.75) {};
		\node [style=none] (140) at (6.75, -6) {};
		\node [style=Z phase dot] (143) at (8.75, -6) {$\frac{\pi}{2}$};
		\node [style=none] (144) at (13, -6) {};
		\node [style=none] (145) at (6.75, -7.25) {};
		\node [style=Z phase dot] (146) at (11, -7.25) {$\frac{\pi}{2}$};
		\node [style=none] (149) at (13, -4.75) {};
		\node [style=none] (150) at (13, -7.25) {};
		\node [style=X dot] (151) at (7.75, -4.75) {};
		\node [style=X dot] (152) at (12, -4.75) {};
		\node [style=none] (153) at (5.75, -6) {$=$};
		\node [style=none] (154) at (16.75, -4.75) {};
		\node [style=none] (155) at (16.75, -6) {};
		\node [style=X phase dot] (158) at (21, -6) {$\pi$};
		\node [style=none] (159) at (23, -6) {};
		\node [style=none] (160) at (16.75, -7.25) {};
		\node [style=X phase dot] (162) at (18.75, -7.25) {$\pi$};
		\node [style=none] (164) at (23, -4.75) {};
		\node [style=none] (165) at (23, -7.25) {};
		\node [style=X phase dot] (166) at (17.75, -4.75) {$\pi$};
		\node [style=X phase dot] (167) at (22, -4.75) {$\pi$};
		\node [style=none] (168) at (14, -6) {$+$};
		\node [style=none] (173) at (4.5, -6) {$...$};
		\node [style=none] (174) at (3.25, -6) {$=$};
		\node [style=none] (175) at (15.75, -6) {$e^{\frac{i\pi}{2}}$};
	\end{pgfonlayer}
	\begin{pgfonlayer}{edgelayer}
		\draw (0.center) to (4);
		\draw (1.center) to (2);
		\draw (2) to (3);
		\draw (3) to (4);
		\draw (3) to (5);
		\draw (5) to (7.center);
		\draw (62.center) to (63);
		\draw (63) to (64);
		\draw (64) to (65);
		\draw (64) to (66);
		\draw (65) to (67.center);
		\draw (66) to (68.center);
		\draw (4) to (65);
		\draw [in=180, out=0] (69.center) to (73);
		\draw (70.center) to (71);
		\draw (71) to (72);
		\draw [bend left, looseness=0.75] (72) to (73);
		\draw (72) to (74);
		\draw (74) to (75.center);
		\draw (76.center) to (77);
		\draw (77) to (78);
		\draw (78) to (80);
		\draw (80) to (82.center);
		\draw (73) to (81.center);
		\draw [bend left] (73) to (78);
		\draw (104.center) to (105);
		\draw (105) to (106);
		\draw (106) to (108);
		\draw (108) to (109.center);
		\draw (110.center) to (111);
		\draw (111) to (112);
		\draw (112) to (113);
		\draw (113) to (115.center);
		\draw (118) to (114.center);
		\draw (117) to (103.center);
		\draw (121.center) to (122);
		\draw (122) to (123);
		\draw (123) to (124);
		\draw (124) to (125.center);
		\draw (126.center) to (127);
		\draw (127) to (128);
		\draw (128) to (129);
		\draw (129) to (131.center);
		\draw (133) to (130.center);
		\draw (132) to (120.center);
		\draw (135) to (106);
		\draw (136) to (112);
		\draw (137) to (123);
		\draw (138) to (128);
		\draw (143) to (144.center);
		\draw (145.center) to (146);
		\draw (152) to (149.center);
		\draw (151) to (139.center);
		\draw (158) to (159.center);
		\draw (167) to (164.center);
		\draw (166) to (154.center);
		\draw (143) to (140.center);
		\draw (146) to (150.center);
		\draw (158) to (155.center);
		\draw (165.center) to (162);
		\draw (162) to (160.center);
	\end{pgfonlayer}
\end{tikzpicture}} \ .
\end{equation}

Note that each term on the RHS of \eqref{eq:picom_grouped} have a $T$-count of $0$. The procedurally optimised cutting algorithm from \cite{Sutcliffe_2024} chooses the order of cutting, in hopes to minimise overall $T$-count.

We shall go through the different sub-routines of magic state cultivation in the next section, before giving explicit stabiliser decompositions by cutting in the later sections. 

\section{Magic state cultivation}
Magic state cultivation is a method to reliably produce logical magic states with logical error rates as low as $10^{-9}$ at the gold standard $0.1\%$ circuit-level noise. The crucial advantage is the low total space-time volume needed to generate such a high quality logical $\ket{T}$ state, at $\sim 10^3$ expected qubit-rounds \cite{gidney2024magicstatecultivationgrowing}. These states are crucial resource states, consumed to perform fault tolerant quantum computation. The authors of \cite{gidney2024magicstatecultivationgrowing} provided a family of two protocols based on the $d=3$ and $d=5$ colour codes; and simulated most of the larger circuits ($d=5$) by replacing the physical $T$ gates entirely with $S$ gates. However, they noticed a slight logical error rate difference between the $S$ and $T$ gates simulations when benchmarked against brute-force state vector simulations for smaller circuit ($d=3$). We wish to find better stabiliser decompositions of the state generated by these circuits, in hopes that this would inspire better simulation methods for the MSC circuit end-to-end.

In this section, for simplicity, we shall use the $d=3$ colour code MSC circuit to illustrate its sub-routine. All the MSC circuit studied in this manuscript is post-selected upon $+1$ measurement results for simple analysis.

In the $d=3$ colour code MSC circuit, a magic state is firstly injected with a variety of methods. In the python code \cite{gidneyy2024cultivationdata} accompanying the paper \cite{gidney2024magicstatecultivationgrowing}, three different methods are available. We choose the $\mathtt{degenerate}$ injection method here:
\begin{equation}
    \label{eq:deg_inj_d_3}
    \scalebox{1}{\input{deg_inj_d_3.tikz}} \ .
\end{equation}

\noindent This sub-circuit has a single $T^{\dagger}$ gate represented by $\scalebox{1}{
\begin{tikzpicture}
    \begin{pgfonlayer}{nodelayer}
        \node [style=Z phase dot] (0) at (-0.00, -0.00) {$\frac{7\pi}{4}$};
        \node [style=none] (1) at (-1.00, -0.00) {};
        \node [style=none] (2) at (1.00, -0.00) {};
    \end{pgfonlayer}
    \begin{pgfonlayer}{edgelayer}
        \draw (0) to (1);
        \draw (0) to (2);
    \end{pgfonlayer}
\end{tikzpicture}
}$. This is followed by a `double checking circuit' that dominates the total $T$-count of the full circuit:
\begin{equation}
    \scalebox{1}{\begin{tikzpicture}
    \begin{pgfonlayer}{nodelayer}
        \node [style=none] (0) at (0.50, -0.50) {};
        \node [style=none] (1) at (13.50, -8.50) {};
        \node [style=Z dot] (2) at (0.50, -6.50) {};
        \node [style=none] (3) at (0.50, -10.50) {};
        \node [style=Z dot] (4) at (11.50, -12.50) {};
        \node [style=Z dot] (5) at (11.50, -4.50) {};
        \node [style=Z dot] (6) at (13.50, -9.50) {};
        \node [style=Z dot] (7) at (11.50, -1.50) {};
        \node [style=X dot] (8) at (3.50, -6.50) {};
        \node [style=Z dot] (9) at (13.50, -12.50) {};
        \node [style=X dot] (10) at (2.50, -11.50) {};
        \node [style=none] (11) at (13.50, -3.50) {};
        \node [style=X dot] (12) at (10.50, -1.50) {};
        \node [style=Z phase dot] (13) at (12.50, -11.50) {$\frac{\pi}{4}$};
        \node [style=Z dot] (14) at (13.50, -4.50) {};
        \node [style=Z phase dot] (15) at (12.50, -10.50) {$\frac{\pi}{4}$};
        \node [style=Z phase dot] (16) at (1.50, -7.50) {$\frac{7\pi}{4}$};
        \node [style=X dot] (17) at (2.50, -7.50) {};
        \node [style=Z dot] (18) at (10.50, -5.50) {};
        \node [style=Z dot] (19) at (2.50, -9.50) {};
        \node [style=Z dot] (20) at (2.50, -6.50) {};
        \node [style=none] (21) at (0.50, -11.50) {};
        \node [style=none] (22) at (13.50, -11.50) {};
        \node [style=X dot] (23) at (5.50, -9.50) {};
        \node [style=Z phase dot] (24) at (1.50, -3.50) {$\frac{7\pi}{4}$};
        \node [style=Z dot] (25) at (11.50, -9.50) {};
        \node [style=Z phase dot] (26) at (12.50, -8.50) {$\frac{\pi}{4}$};
        \node [style=Z phase dot] (27) at (1.50, -0.50) {$\frac{7\pi}{4}$};
        \node [style=Z dot] (28) at (6.50, -5.50) {};
        \node [style=Z dot] (29) at (0.50, -4.50) {};
        \node [style=X dot] (30) at (9.50, -3.50) {};
        \node [style=X dot] (31) at (11.50, -7.50) {};
        \node [style=X dot] (32) at (4.50, -3.50) {};
        \node [style=X dot] (33) at (9.50, -10.50) {};
        \node [style=Z dot] (34) at (3.50, -3.50) {};
        \node [style=X dot] (35) at (11.50, -11.50) {};
        \node [style=Z phase dot] (36) at (1.50, -5.50) {$\frac{7\pi}{4}$};
        \node [style=Z dot] (37) at (9.50, -5.50) {};
        \node [style=X dot] (38) at (11.50, -0.50) {};
        \node [style=Z dot] (39) at (4.50, -9.50) {};
        \node [style=none] (40) at (0.50, -3.50) {};
        \node [style=Z dot] (41) at (5.50, -5.50) {};
        \node [style=Z dot] (42) at (0.50, -2.50) {};
        \node [style=none] (43) at (0.50, -5.50) {};
        \node [style=Z dot] (44) at (2.50, -1.50) {};
        \node [style=X dot] (45) at (11.50, -8.50) {};
        \node [style=X dot] (46) at (10.50, -12.50) {};
        \node [style=X dot] (47) at (11.50, -3.50) {};
        \node [style=Z dot] (48) at (3.50, -5.50) {};
        \node [style=Z phase dot] (49) at (1.50, -10.50) {$\frac{7\pi}{4}$};
        \node [style=X dot] (50) at (2.50, -3.50) {};
        \node [style=Z phase dot] (51) at (12.50, -5.50) {$\frac{\pi}{4}$};
        \node [style=none] (52) at (13.50, -7.50) {};
        \node [style=X dot] (53) at (2.50, -0.50) {};
        \node [style=Z phase dot] (54) at (1.50, -11.50) {$\frac{7\pi}{4}$};
        \node [style=Z dot] (55) at (10.50, -9.50) {};
        \node [style=Z dot] (56) at (10.50, -3.50) {};
        \node [style=Z dot] (57) at (2.50, -2.50) {};
        \node [style=Z phase dot] (58) at (12.50, -7.50) {$\frac{\pi}{4}$};
        \node [style=Z dot] (59) at (0.50, -9.50) {};
        \node [style=none] (60) at (0.50, -8.50) {};
        \node [style=Z phase dot] (61) at (12.50, -0.50) {$\frac{\pi}{4}$};
        \node [style=Z dot] (62) at (9.50, -9.50) {};
        \node [style=Z dot] (63) at (7.50, -5.50) {};
        \node [style=X dot] (64) at (3.50, -1.50) {};
        \node [style=Z dot] (65) at (11.50, -2.50) {};
        \node [style=X dot] (66) at (4.50, -10.50) {};
        \node [style=Z dot] (67) at (13.50, -6.50) {};
        \node [style=X dot] (68) at (8.50, -9.50) {};
        \node [style=X dot] (69) at (2.50, -8.50) {};
        \node [style=Z dot] (70) at (0.50, -1.50) {};
        \node [style=Z dot] (71) at (13.50, -1.50) {};
        \node [style=Z dot] (72) at (11.50, -6.50) {};
        \node [style=Z dot] (73) at (4.50, -5.50) {};
        \node [style=Z dot] (74) at (3.50, -9.50) {};
        \node [style=none] (75) at (13.50, -5.50) {};
        \node [style=Z dot] (76) at (13.50, -2.50) {};
        \node [style=X dot] (77) at (2.50, -5.50) {};
        \node [style=Z dot] (78) at (0.50, -12.50) {};
        \node [style=X dot] (79) at (3.50, -12.50) {};
        \node [style=Z dot] (80) at (8.50, -5.50) {};
        \node [style=Z phase dot] (81) at (1.50, -8.50) {$\frac{7\pi}{4}$};
        \node [style=Z phase dot] (82) at (12.50, -3.50) {$\frac{\pi}{4}$};
        \node [style=none] (83) at (13.50, -0.50) {};
        \node [style=Z dot] (84) at (2.50, -12.50) {};
        \node [style=none] (85) at (0.50, -7.50) {};
        \node [style=X dot] (86) at (10.50, -6.50) {};
        \node [style=none] (87) at (13.50, -10.50) {};
        \node [style=Z dot] (88) at (2.50, -4.50) {};
        \node [style=X dot] (89) at (11.50, -5.50) {};
    \end{pgfonlayer}
    \begin{pgfonlayer}{edgelayer}
        \draw (0) to (27);
        \draw (1) to (26);
        \draw (2) to (20);
        \draw (3) to (49);
        \draw (4) to (9);
        \draw (4) to (35);
        \draw (4) to (46);
        \draw (5) to (14);
        \draw (5) to (88);
        \draw (5) to (89);
        \draw (6) to (25);
        \draw (7) to (71);
        \draw (7) to (12);
        \draw (7) to (38);
        \draw (8) to (48);
        \draw (8) to (86);
        \draw (8) to (20);
        \draw (10) to (84);
        \draw (10) to (35);
        \draw (10) to (54);
        \draw (11) to (82);
        \draw (12) to (56);
        \draw (12) to (64);
        \draw (13) to (22);
        \draw (13) to (35);
        \draw (15) to (87);
        \draw (15) to (33);
        \draw (16) to (17);
        \draw (16) to (85);
        \draw (17) to (20);
        \draw (17) to (31);
        \draw (18) to (89);
        \draw (18) to (37);
        \draw (18) to (86);
        \draw (19) to (74);
        \draw (19) to (59);
        \draw (19) to (69);
        \draw (21) to (54);
        \draw (23) to (41);
        \draw (23) to (68);
        \draw (23) to (39);
        \draw (24) to (50);
        \draw (24) to (40);
        \draw (25) to (45);
        \draw (25) to (55);
        \draw (26) to (45);
        \draw (27) to (53);
        \draw (28) to (41);
        \draw (29) to (88);
        \draw (30) to (37);
        \draw (30) to (56);
        \draw (30) to (32);
        \draw (31) to (72);
        \draw (31) to (58);
        \draw (32) to (73);
        \draw (32) to (34);
        \draw (33) to (62);
        \draw (33) to (66);
        \draw (34) to (50);
        \draw (34) to (64);
        \draw (36) to (77);
        \draw (36) to (43);
        \draw (37) to (80);
        \draw (38) to (61);
        \draw (38) to (53);
        \draw (39) to (66);
        \draw (39) to (74);
        \draw (41) to (73);
        \draw (42) to (57);
        \draw (44) to (64);
        \draw (44) to (53);
        \draw (44) to (70);
        \draw (45) to (69);
        \draw (46) to (55);
        \draw (46) to (79);
        \draw (47) to (65);
        \draw (47) to (82);
        \draw (47) to (56);
        \draw (48) to (73);
        \draw (48) to (77);
        \draw (49) to (66);
        \draw (50) to (57);
        \draw (51) to (75);
        \draw (51) to (89);
        \draw (52) to (58);
        \draw (55) to (62);
        \draw (57) to (65);
        \draw (60) to (81);
        \draw (61) to (83);
        \draw (62) to (68);
        \draw (63) to (80);
        \draw (65) to (76);
        \draw (67) to (72);
        \draw (68) to (80);
        \draw (69) to (81);
        \draw (72) to (86);
        \draw (74) to (79);
        \draw (77) to (88);
        \draw (78) to (84);
        \draw (79) to (84);
    \end{pgfonlayer}
\end{tikzpicture}} \ .
\end{equation}

\noindent The $d=3$ MSC circuit has $14$ $T$ gates in the double checking sub-routine and $15$ in total:
\begin{equation}
    \label{eq:MSC_d_3_full}
    \scalebox{0.75}{\includegraphics{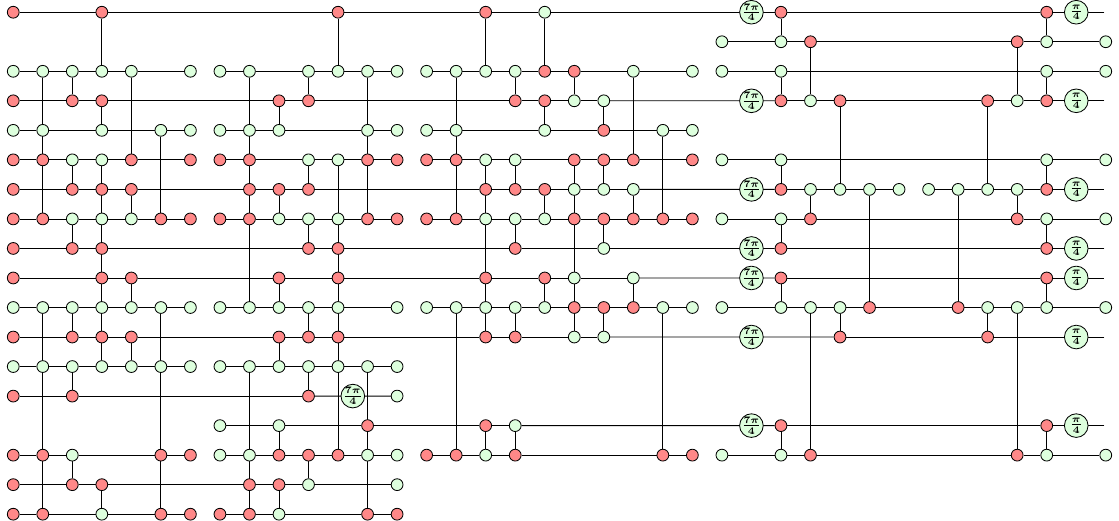}} \ .
\end{equation}

\noindent The final $7$-legged encoded state can then be `grafted' onto a surface code in the escape stage (which we don't include here). We will look at the ZX-diagram/state specified by \eqref{eq:MSC_d_3_full} and find its cutting stabiliser decomposition in the next section.

\section{Stabiliser decomposition of the $d=3$ colour code MSC circuit}
In order to use the cutting stabiliser decomposition on the circuit from \eqref{eq:MSC_d_3_full}, we first have to simplify this ZX-diagram by fusing spiders together:
\begin{equation}
    \label{eq:MSC_d_3_full_simp}
    \scalebox{0.6}{\includegraphics{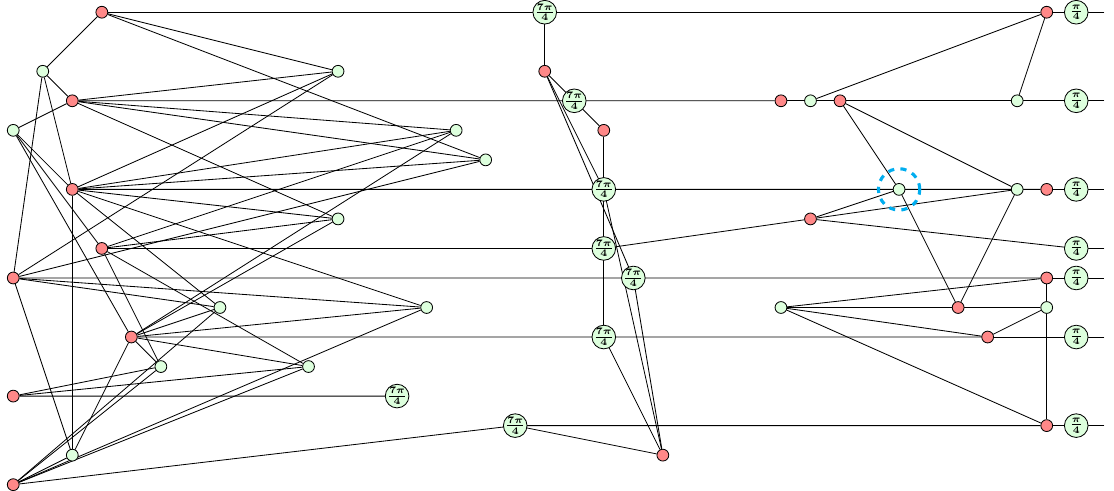}} \ .
\end{equation}

\noindent For the first node cutting, we used the procedurally optimised cutting algorithm from \cite{Sutcliffe_2024,Sutcliffe2025thesis} to choose the spider inside the dashed {\color{cyan}cyan} coloured circle node in \ref{eq:MSC_d_3_full_simp} to be cut. This cutting decomposed the ZX-diagram from \eqref{eq:MSC_d_3_full_simp} into a sum of $2$ terms as shown in \eqref{eq:d3_cut1}:


\begin{equation}
\centering
\label{eq:d3_cut1}
\begin{tikzpicture}[every node/.style={anchor=center},baseline=(A1.center)]
  \node (A1) {\scalebox{0.3}{\includegraphics{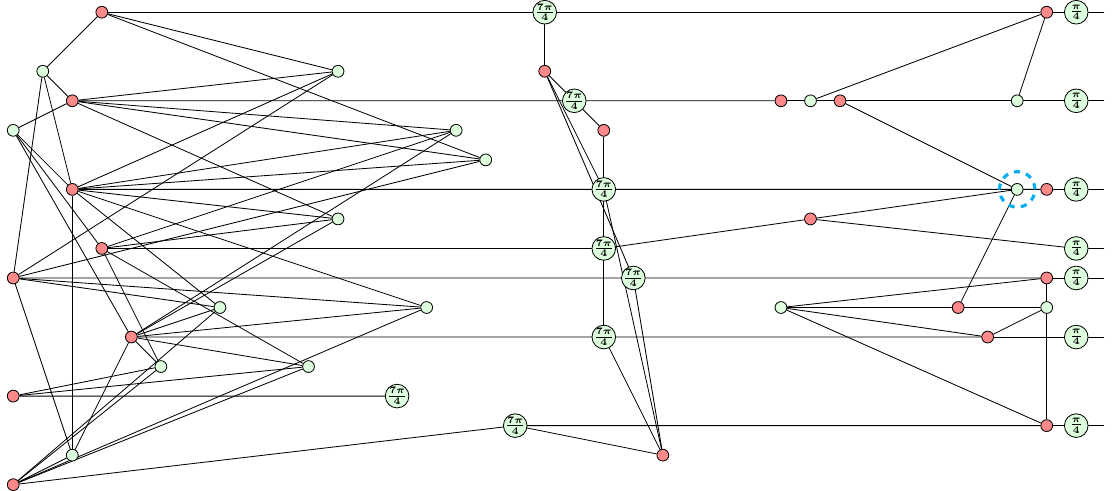}}};
  \node[right=0.3cm of A1] (plus1) {$+$};
  \node[right=0.3cm of plus1] (A2) {\scalebox{0.3}{\includegraphics{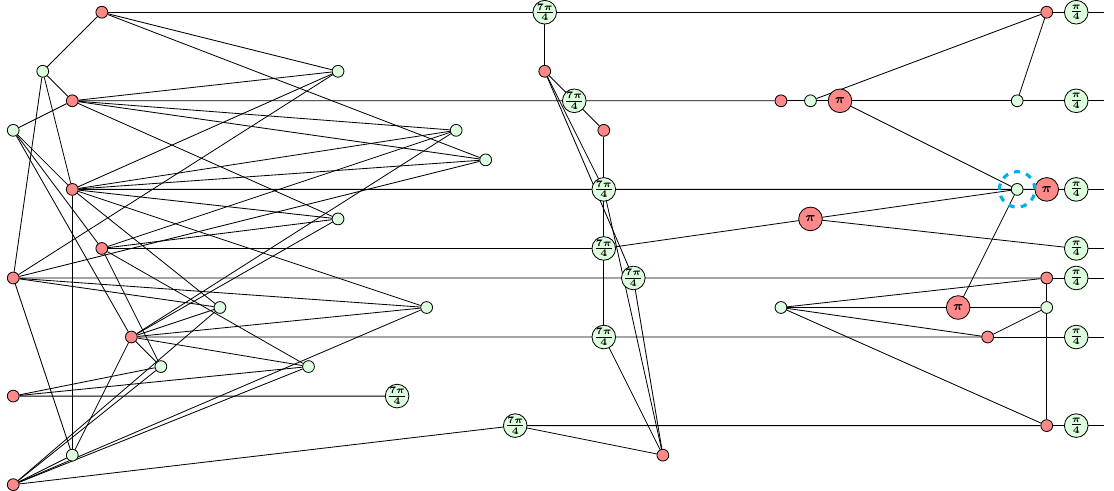}}};
\end{tikzpicture}
\end{equation}

\noindent We then further choose the spider inside the dashed {\color{cyan}cyan} coloured circle node in \ref{eq:d3_cut1} for the second graph cutting on each of the two terms. After the second cutting we have 4 terms as shown in \eqref{eq:cut2_b4_simp}.
\begin{equation}
\centering
\label{eq:cut2_b4_simp}
\begin{tikzpicture}[every node/.style={anchor=center},baseline=(A3.center)]
  \node (A1) {\scalebox{0.3}{\includegraphics{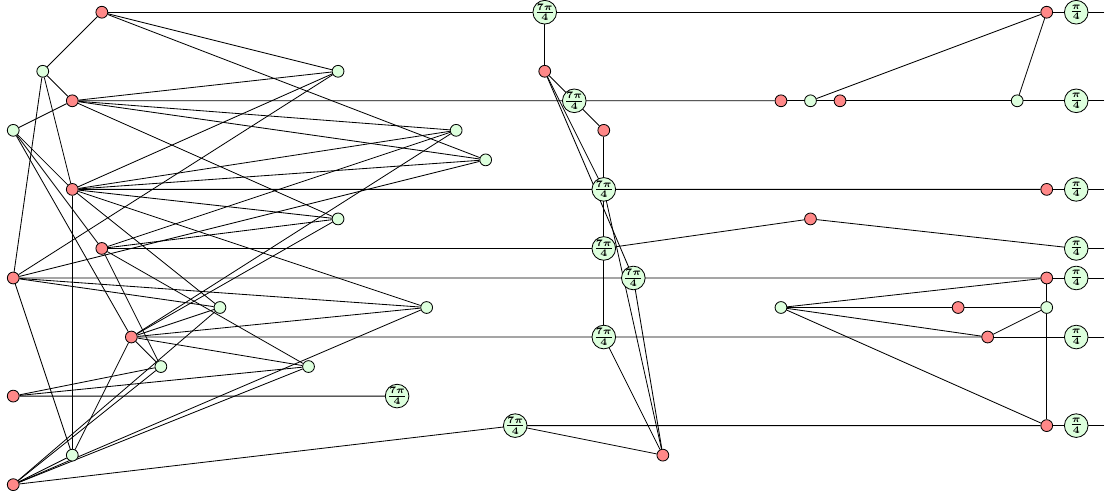}}};
  \node[right=0.3cm of A1] (plus1) {$+$};
  \node[right=0.3cm of plus1] (A2) {\scalebox{0.3}{\includegraphics{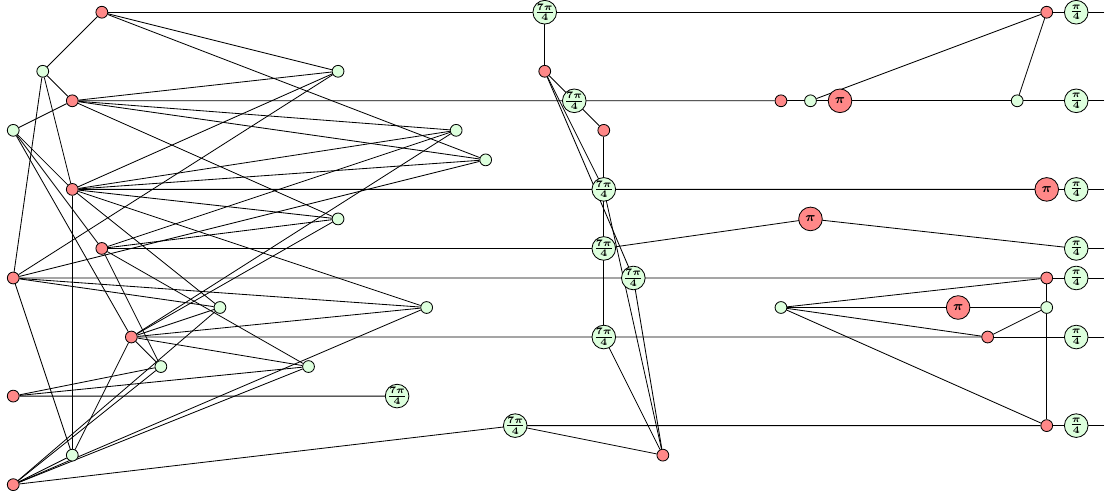}}};
  \node[below=0.3cm of A1] (A3) {\scalebox{0.3}{\includegraphics{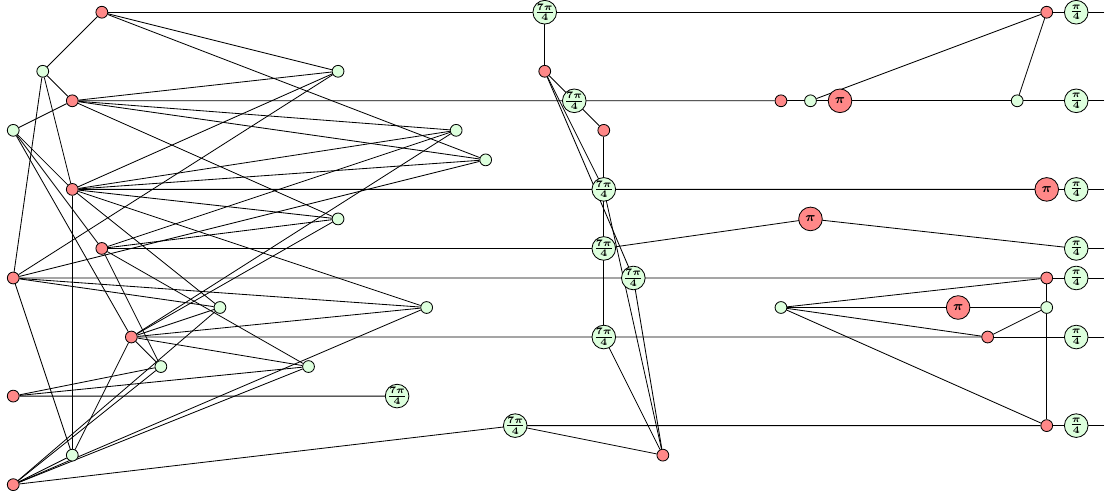}}};
  \node[right=0.3cm of A3] (plus2) {$+$};
  \node[right=0.3cm of plus2] (A4) {\scalebox{0.3}{\includegraphics{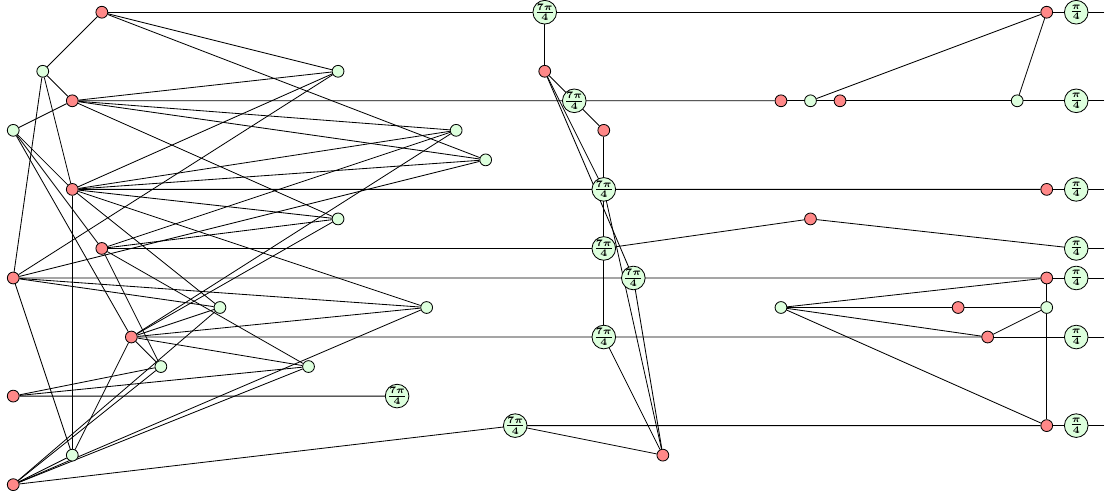}}};
  \node[left=0.3cm of A3] (plus3) {$+$};
\end{tikzpicture}
\end{equation}

\noindent We notice the ZX-diagrams in the top left and bottom right of \eqref{eq:cut2_b4_simp} are identical, hence they equate to a single term. Similarly the bottom left and top right ZX-diagrams are also identical, hence they also can be combined into a single term. After reducing the total number of terms by half, we arrive at \eqref{eq:cut2_b4_simp_red}.
\begin{equation}
\centering
\label{eq:cut2_b4_simp_red}
\begin{tikzpicture}[every node/.style={anchor=center},baseline=(A1.center)]
  \node (A1) {\scalebox{0.3}{\includegraphics{cut2_term0.pdf}}};
  \node[right=0.3cm of A1] (plus1) {$+$};
  \node[right=0.3cm of plus1] (A2) {\scalebox{0.3}{\includegraphics{cut2_term1.pdf}}};
\end{tikzpicture}
\end{equation}

\noindent The real magic comes when you simplify\footnote{Effectively we performed $\mathtt{zx.full\_reduce(g)}$ in $\mathtt{pyzx}$.} each ZX-diagram term separately in \eqref{eq:cut2_b4_simp_red}. Now each term has a $T$-count of one!
\begin{equation}
\centering
\label{eq:cut2_after_simp}
\begin{tikzpicture}[every node/.style={anchor=center},baseline=(A1.center)]
  \node (A1) {\scalebox{0.75}{\begin{tikzpicture}
    \begin{pgfonlayer}{nodelayer}
        \node [style=none] (1) at (24.00, -4.00) {};
        \node [style=Z dot] (2) at (18.00, -0.50) {};
        \node [style=Z phase dot] (3) at (12.00, 1.00) {$\frac{7\pi}{4}$};
        \node [style=none] (4) at (24.00, -6.00) {};
        \node [style=none] (5) at (24.00, -1.00) {};
        \node [style=none] (6) at (24.00, -0.00) {};
        \node [style=Z dot] (7) at (23.00, -3.00) {};
        \node [style=none] (8) at (24.00, -5.00) {};
        \node [style=Z dot] (9) at (23.00, -2.00) {};
        \node [style=Z dot] (10) at (12.00, -2.00) {};
        \node [style=Z dot] (11) at (19.00, -1.25) {};
        \node [style=Z dot] (12) at (23.00, -6.00) {};
        \node [style=Z dot] (13) at (13.00, -0.00) {};
        \node [style=Z dot] (14) at (23.00, -1.00) {};
        \node [style=none] (15) at (24.00, -2.00) {};
        \node [style=Z dot] (16) at (23.00, -4.00) {};
        \node [style=Z dot] (17) at (20.75, -2.00) {};
        \node [style=Z dot] (18) at (12.00, -6.00) {};
        \node [style=none] (19) at (24.00, -3.00) {};
        \node [style=Z dot] (20) at (14.25, -4.00) {};
        \node [style=Z dot] (21) at (23.00, -5.00) {};
        \node [style=Z dot] (22) at (23.00, -0.00) {};
        \node [style=Z dot] (23) at (15.25, -2.25) {};
    \end{pgfonlayer}
    \begin{pgfonlayer}{edgelayer}
        \draw (1) to (16);
        \draw [style=hadamard edge] (2) to (23);
        \draw [style=hadamard edge] (2) to (11);
        \draw [style=hadamard edge] (2) to (9);
        \draw [style=hadamard edge] (2) to (10);
        \draw (3) to (10);
        \draw (4) to (12);
        \draw (5) to (14);
        \draw (6) to (22);
        \draw (7) to (19);
        \draw [style=hadamard edge] (7) to (17);
        \draw (8) to (21);
        \draw (9) to (15);
        \draw [style=hadamard edge] (10) to (13);
        \draw [style=hadamard edge] (10) to (18);
        \draw (11) to (14);
        \draw [style=hadamard edge] (11) to (17);
        \draw [style=hadamard edge] (11) to (18);
        \draw (12) to (23);
        \draw [style=hadamard edge] (13) to (20);
        \draw [style=hadamard edge] (13) to (23);
        \draw [style=hadamard edge] (13) to (22);
        \draw (16) to (20);
        \draw [style=hadamard edge] (17) to (20);
        \draw [style=hadamard edge] (17) to (23);
        \draw [style=hadamard edge] (18) to (20);
        \draw [style=hadamard edge] (18) to (21);
    \end{pgfonlayer}
\end{tikzpicture}}};
  \node[right=0.3cm of A1] (plus1) {$+$};
  \node[right=0.3cm of plus1] (A2) {\scalebox{0.75}{\begin{tikzpicture}
    \begin{pgfonlayer}{nodelayer}
        \node [style=Z dot] (0) at (27.50, -0.00) {};
        \node [style=Z dot] (1) at (31.50, -2.25) {};
        \node [style=Z dot] (2) at (29.50, 0.25) {};
        \node [style=Z dot] (3) at (34.50, -0.00) {};
        \node [style=Z dot] (4) at (30.25, -4.25) {};
        \node [style=Z phase dot] (5) at (27.50, -6.00) {$\frac{3\pi}{2}$};
        \node [style=Z phase dot] (6) at (31.50, -4.25) {$\frac{3\pi}{2}$};
        \node [style=Z phase dot] (7) at (31.50, -1.25) {$\frac{3\pi}{2}$};
        \node [style=Z phase dot] (8) at (27.50, 1.00) {$\frac{\pi}{4}$};
        \node [style=Z dot] (9) at (37.00, 0.00) {};
        \node [style=Z dot] (10) at (37.00, -1.00) {};
        \node [style=Z dot] (11) at (37.00, -2.00) {};
        \node [style=Z dot] (12) at (37.00, -3.00) {};
        \node [style=Z dot] (13) at (37.00, -4.00) {};
        \node [style=Z dot] (14) at (37.00, -5.00) {};
        \node [style=Z dot] (15) at (37.00, -6.00) {};
        \node [style=none] (16) at (38.00, 0.00) {};
        \node [style=none] (17) at (38.00, -1.00) {};
        \node [style=none] (18) at (38.00, -2.00) {};
        \node [style=none] (19) at (38.00, -3.00) {};
        \node [style=none] (20) at (38.00, -4.00) {};
        \node [style=none] (21) at (38.00, -5.00) {};
        \node [style=none] (22) at (38.00, -6.00) {};
    \end{pgfonlayer}
    \begin{pgfonlayer}{edgelayer}
        \draw [style=hadamard edge] (0) to (8);
        \draw [style=hadamard edge] (0) to (6);
        \draw [style=hadamard edge] (0) to (7);
        \draw [style=hadamard edge] (0) to (5);
        \draw [style=hadamard edge] (1) to (4);
        \draw [style=hadamard edge] (1) to (2);
        \draw [style=hadamard edge] (1) to (3);
        \draw (1) to (13);
        \draw [style=hadamard edge] (2) to (7);
        \draw [style=hadamard edge] (2) to (5);
        \draw [style=hadamard edge] (2) to (10);
        \draw [style=hadamard edge] (3) to (6);
        \draw [style=hadamard edge] (3) to (7);
        \draw [style=hadamard edge] (3) to (9);
        \draw [style=hadamard edge] (4) to (5);
        \draw [style=hadamard edge] (4) to (6);
        \draw [style=hadamard edge] (4) to (11);
        \draw [style=hadamard edge] (5) to (7);
        \draw [style=hadamard edge] (5) to (6);
        \draw (5) to (15);
        \draw [style=hadamard edge] (6) to (7);
        \draw (6) to (14);
        \draw (7) to (12);
        \draw (9) to (16);
        \draw (10) to (17);
        \draw (11) to (18);
        \draw (12) to (19);
        \draw (13) to (20);
        \draw (14) to (21);
        \draw (15) to (22);
    \end{pgfonlayer}
\end{tikzpicture}}};
\end{tikzpicture}
\end{equation}

\noindent We can further decompose each $\scalebox{1}{
\begin{tikzpicture}
    \begin{pgfonlayer}{nodelayer}
        \node [style=Z phase dot] (0) at (-2.00, -0.00) {$\frac{\pi}{4}$};
        \node [style=none] (1) at (-1.00, -0.00) {};
    \end{pgfonlayer}
    \begin{pgfonlayer}{edgelayer}
        \draw (0) to (1);
    \end{pgfonlayer}
\end{tikzpicture}
}$, $\scalebox{1}{
\begin{tikzpicture}
    \begin{pgfonlayer}{nodelayer}
        \node [style=Z phase dot] (0) at (-2.00, -0.00) {$\frac{7\pi}{4}$};
        \node [style=none] (1) at (-1.00, -0.00) {};
    \end{pgfonlayer}
    \begin{pgfonlayer}{edgelayer}
        \draw (0) to (1);
    \end{pgfonlayer}
\end{tikzpicture}
}$ spider in \eqref{eq:cut2_after_simp} into a superposition of $2$ terms using the following \cite{Kiss2022}:
\begin{equation}
    \label{eq:single_magic}
    \begin{split}
        \begin{tikzpicture}[scale=2]
	\begin{pgfonlayer}{nodelayer}
		\node [style=Z phase dot] (0) at (0, -0.012) {$\frac\pi4$};
		\node [style=none] (1) at (0.525, -0.013) {};
	\end{pgfonlayer}
	\begin{pgfonlayer}{edgelayer}
		\draw (1.center) to (0);
	\end{pgfonlayer}
\end{tikzpicture}
 & = \frac1{\sqrt2}\left(\begin{tikzpicture}
	\begin{pgfonlayer}{nodelayer}
		\node [style=Z dot] (0) at (-1, 0) {};
		\node [style=Z dot] (1) at (0, 0) {};
		\node [style=none] (2) at (1, 0) {};
	\end{pgfonlayer}
	\begin{pgfonlayer}{edgelayer}
		\draw (1) to (2.center);
		\draw [style=hadamard edge] (0) to (1);
	\end{pgfonlayer}
\end{tikzpicture}
+e^{i\frac\pi4}\begin{tikzpicture}
	\begin{pgfonlayer}{nodelayer}
		\node [style=Z phase dot] (0) at (-1, 0) {$\pi$};
		\node [style=Z dot] (1) at (0.25, 0) {};
		\node [style=none] (2) at (1.25, 0) {};
	\end{pgfonlayer}
	\begin{pgfonlayer}{edgelayer}
		\draw (1) to (2.center);
		\draw [style=hadamard edge] (0) to (1);
	\end{pgfonlayer}
\end{tikzpicture}
\right) \ , \\ 
        \begin{tikzpicture}[scale=2]
	\begin{pgfonlayer}{nodelayer}
		\node [style=Z phase dot] (0) at (0, -0.012) {$\frac{7\pi}{4}$};
		\node [style=none] (1) at (0.525, -0.013) {};
	\end{pgfonlayer}
	\begin{pgfonlayer}{edgelayer}
		\draw (1.center) to (0);
	\end{pgfonlayer}
\end{tikzpicture}
 & = \frac1{\sqrt2}\left(\begin{tikzpicture}
	\begin{pgfonlayer}{nodelayer}
		\node [style=Z dot] (0) at (-1, 0) {};
		\node [style=Z dot] (1) at (0, 0) {};
		\node [style=none] (2) at (1, 0) {};
	\end{pgfonlayer}
	\begin{pgfonlayer}{edgelayer}
		\draw (1) to (2.center);
		\draw [style=hadamard edge] (0) to (1);
	\end{pgfonlayer}
\end{tikzpicture}
+e^{-i\frac\pi4}\begin{tikzpicture}
	\begin{pgfonlayer}{nodelayer}
		\node [style=Z phase dot] (0) at (-1, 0) {$\pi$};
		\node [style=Z dot] (1) at (0.25, 0) {};
		\node [style=none] (2) at (1.25, 0) {};
	\end{pgfonlayer}
	\begin{pgfonlayer}{edgelayer}
		\draw (1) to (2.center);
		\draw [style=hadamard edge] (0) to (1);
	\end{pgfonlayer}
\end{tikzpicture}
\right) \ .
            \end{split}
\end{equation}

\noindent Hence, the the resultant state from \eqref{eq:MSC_d_3_full} are can be represented as a sum of $2\times 2 = 4$ total Clifford ZX-diagram. We will move onto the $d=5$ variant of the MSC circuit in the next section, with some details omitted, but the computation is largely similar.

\section{Stabiliser decomposition of the $d=5$ colour code MSC circuit}
The $d=5$ colour code MSC circuit\footnote{Similarly to the previous circuit, we choose the $\mathtt{degenerate}$ state injection and post-select all measurement results at $+1$ for consistency and simplicity.}, as shown in \eqref{eq:MSC_d_5_full}, can be seen as the resultant state of $d=3$ variant (the $d=3$ variant lies inside the black dashed box in \eqref{eq:MSC_d_5_full}) followed by a series of circuits and then a larger double checking circuit containing a $T$-count of $38$. 
\begin{equation}
    \label{eq:MSC_d_5_full}
    \scalebox{0.27}{\includegraphics{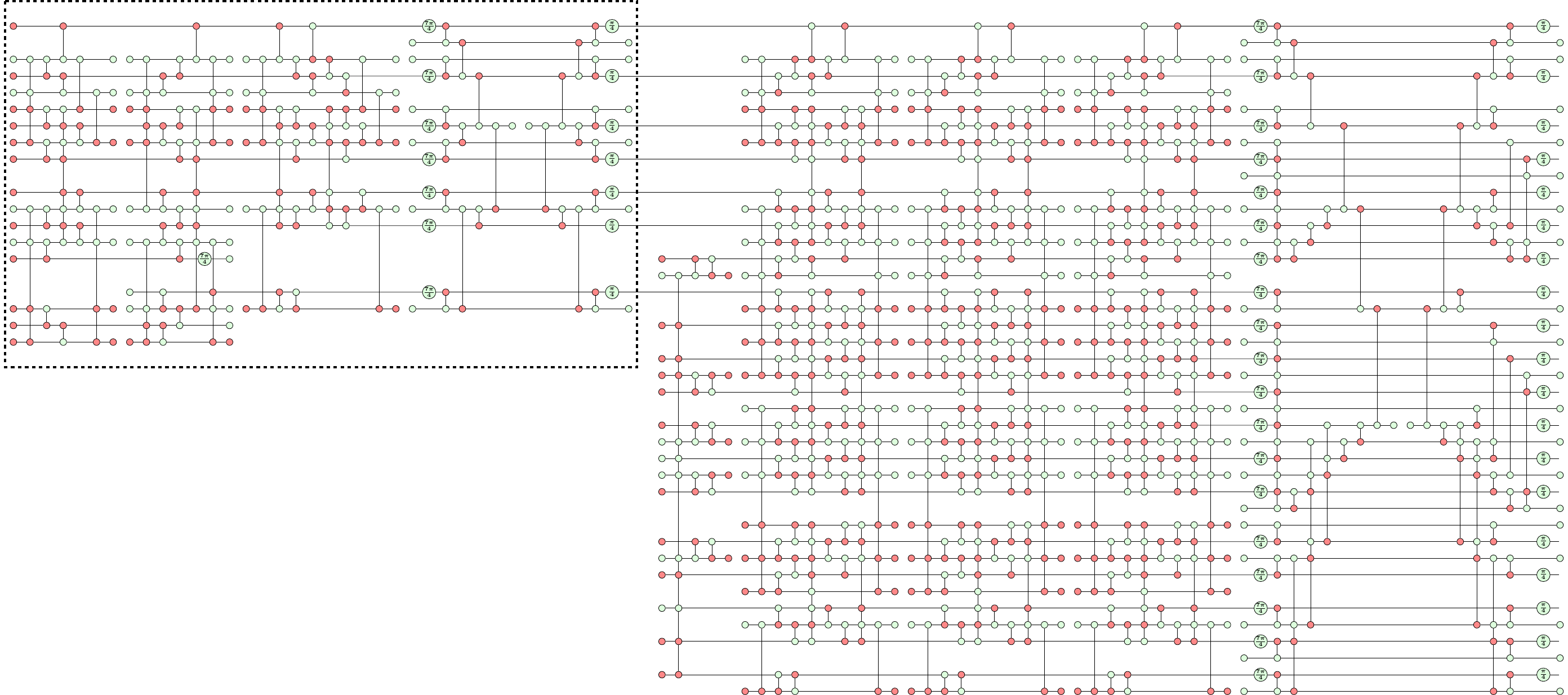}}
\end{equation}

\noindent This full $d=5$ colour code MSC circuit has a $T$-count of $53$. 

There are two different strategies we could employ to obtain stabiliser decompositions to this ZX-diagram. Firstly, we can just apply the standard procedurally optimised cutting algorithm to find a cutting stabiliser decomposition. Alternatively, we can use the stabiliser decomposition results obtain in the $d=3$ variant, and replace the according sub-circuit of the $d=5$ colour code MSC circuit with the already `cut' stabiliser decompositions, before feeding that again into the procedurally optimised cutting algorithm.

\subsection{Naive procedurally optimised cutting}
We can first try to use the procedurally optimised cutting algorithm. We simplify the ZX-diagram from \eqref{eq:MSC_d_5_full}. Note that there are lone $\scalebox{1}{\begin{tikzpicture}[scale=2]
	\begin{pgfonlayer}{nodelayer}
		\node [style=Z dot] (0) at (0, -0.012) {};
	\end{pgfonlayer}
	\begin{pgfonlayer}{edgelayer}
	\end{pgfonlayer}
\end{tikzpicture}
}$/$\scalebox{1}{\begin{tikzpicture}[scale=2]
	\begin{pgfonlayer}{nodelayer}
		\node [style=X dot] (0) at (0, -0.012) {};
	\end{pgfonlayer}
	\begin{pgfonlayer}{edgelayer}
	\end{pgfonlayer}
\end{tikzpicture}
}$ spiders that are results from the simplifications of flag-like checks $\Big(\scalebox{1}{\begin{tikzpicture}
    \begin{pgfonlayer}{nodelayer}
        \node [style=none] (1) at (1.00, 0.50) {};
        \node [style=Z dot] (2) at (-2.00, -0.50) {};
        \node [style=X dot] (3) at (-2.00, 0.50) {};
        \node [style=X dot] (5) at (0.00, 0.50) {};
        \node [style=Z dot] (6) at (0.00, -0.50) {};
        \node [style=none] (7) at (-3.00, 0.50) {};
    \end{pgfonlayer}
    \begin{pgfonlayer}{edgelayer}
        \draw (1) to (5);
        \draw (2) to (3);
        \draw (2) to (6);
        \draw (3) to (5);
        \draw (3) to (7);
        \draw (5) to (6);
    \end{pgfonlayer}
\end{tikzpicture}} = \scalebox{1}{\begin{tikzpicture}
    \begin{pgfonlayer}{nodelayer}
        \node [style=Z dot] (0) at (0.00, -0.50) {};
        \node [style=none] (1) at (-1.00, 0.50) {};
        \node [style=none] (2) at (1.00, 0.50) {};
        \node [style=X dot] (3) at (0.00, 0.50) {};
    \end{pgfonlayer}
    \begin{pgfonlayer}{edgelayer}
        \draw [out=180, in=250, looseness=1.2] (0) to (3);
        \draw (0) to (3);
        \draw (1) to (3);
        \draw (2) to (3);
    \end{pgfonlayer}
\end{tikzpicture}} = \scalebox{1}{\begin{tikzpicture}
    \begin{pgfonlayer}{nodelayer}
        \node [style=Z dot] (0) at (0.00, -0.50) {};
        \node [style=none] (1) at (-1.00, 0.50) {};
        \node [style=none] (2) at (1.00, 0.50) {};
        \node [style=X dot] (3) at (0.00, 0.50) {};
    \end{pgfonlayer}
    \begin{pgfonlayer}{edgelayer}
        \draw (1) to (3);
        \draw (2) to (3);
    \end{pgfonlayer}
\end{tikzpicture}}\Big)$ in the MSC circuits \eqref{eq:MSC_d_5_full}.
\begin{equation}
    \label{eq:MSC_d_5_full_simp}
    \scalebox{0.27}{\includegraphics{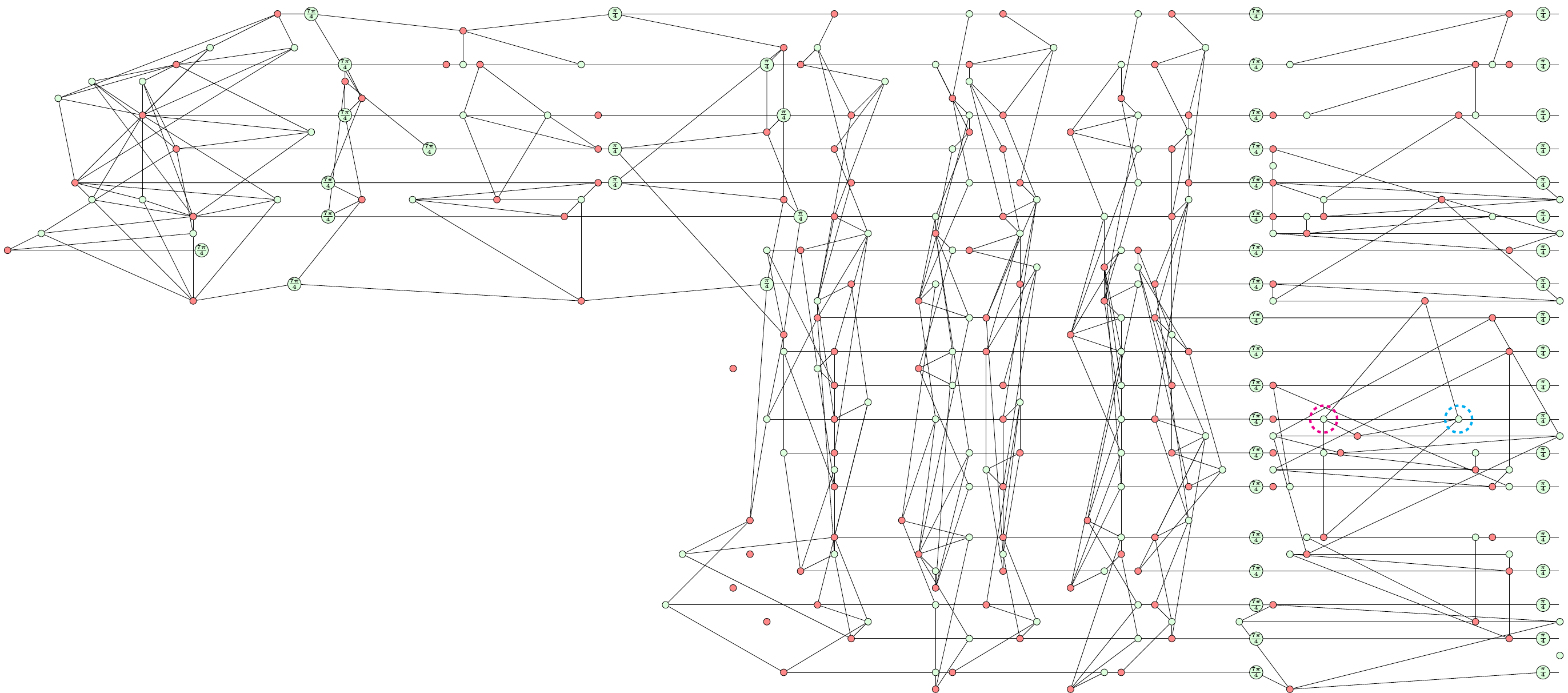}}
\end{equation}

\noindent We make a first cut with the spider inside the dashed {\color{cyan}cyan} coloured circle, followed by a second cut with the spider inside the dashed {\color{magenta}magenta} coloured circle. After $2$ cuts, we will have $2^2=4$ terms. We simplify these $4$ terms individually via ZX-rewrites and it turns out each individual term has identical $T$-count of $15$:
\begin{equation}
\centering
\label{eq:naive_d5_term}
\begin{tikzpicture}[every node/.style={anchor=center},baseline=(A3.center)]
  \node (A1) {\scalebox{0.2}{\includegraphics{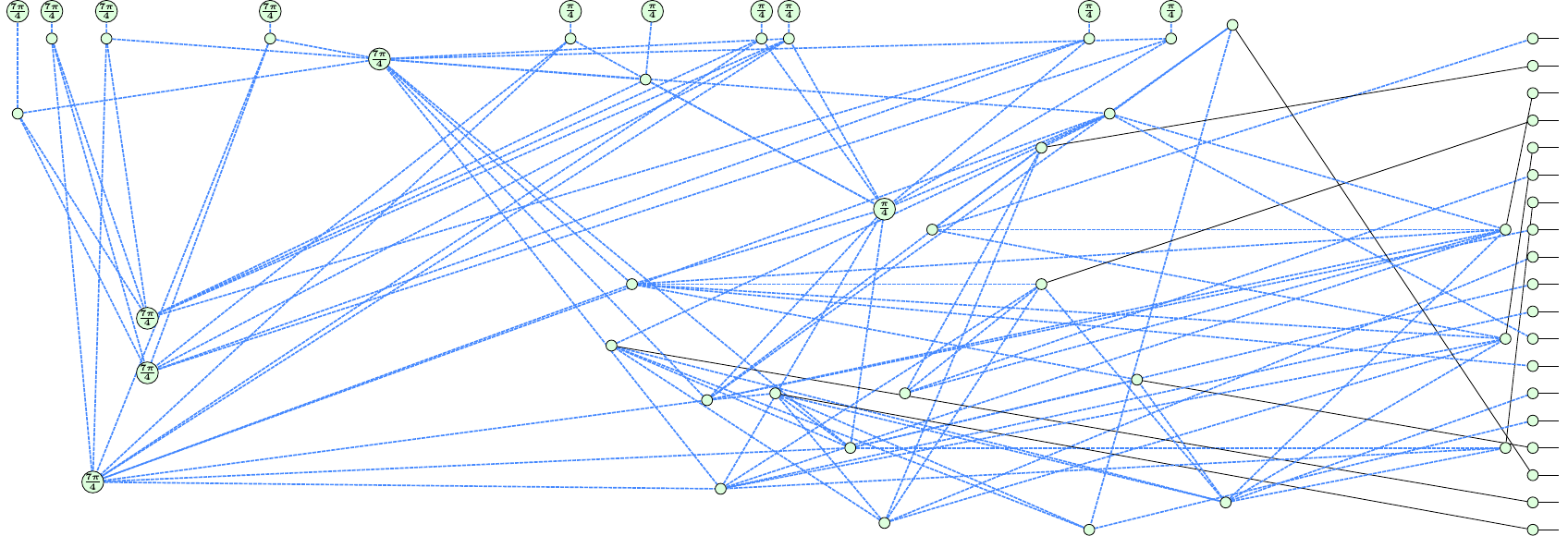}}};
  \node[right=0.3cm of A1] (plus1) {$+$};
  \node[right=0.3cm of plus1] (A2) {\scalebox{0.2}{\includegraphics{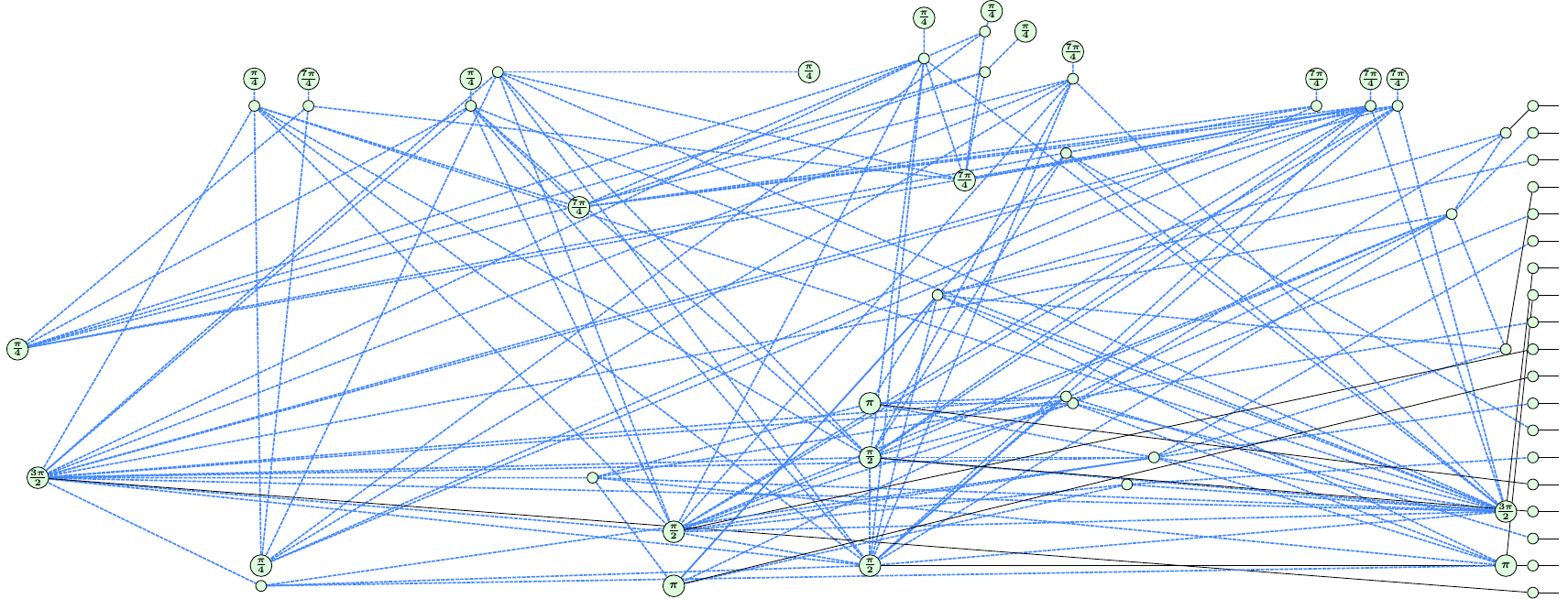}}};
  \node[below=0.3cm of A1] (A3) {\scalebox{0.2}{\includegraphics{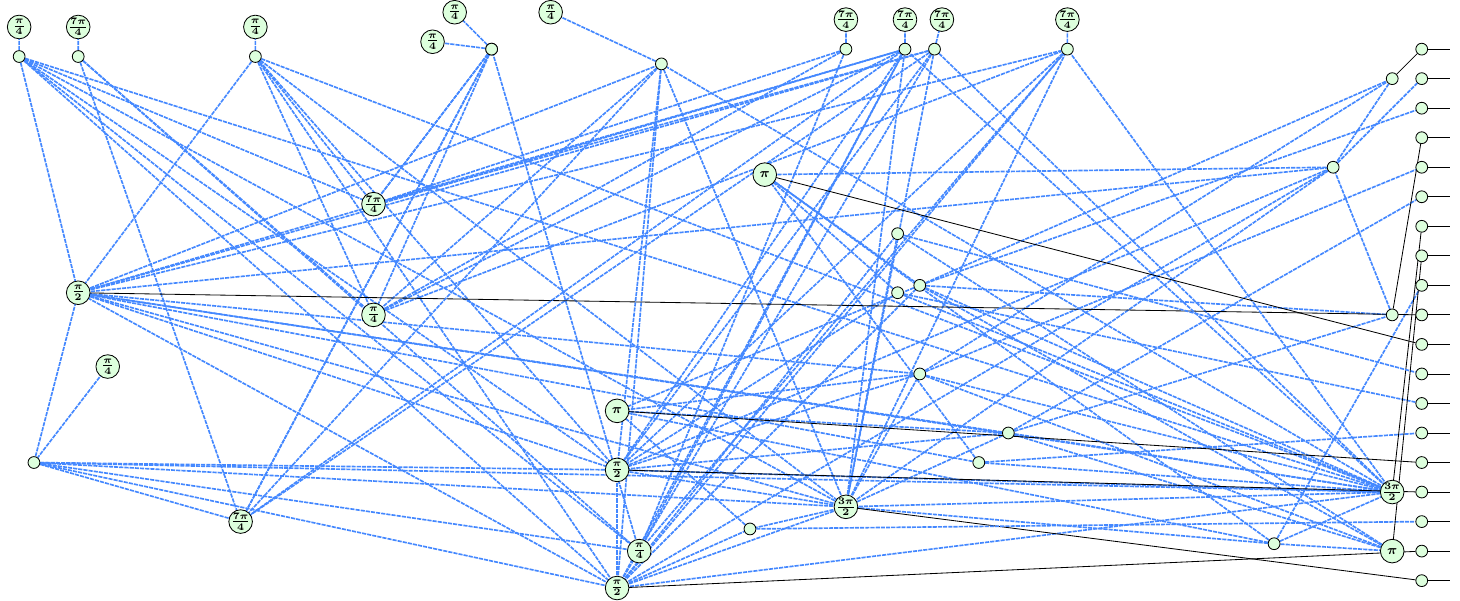}}};
  \node[right=0.3cm of A3] (plus2) {$+$};
  \node[right=0.3cm of plus2] (A4) {\scalebox{0.2}{\includegraphics{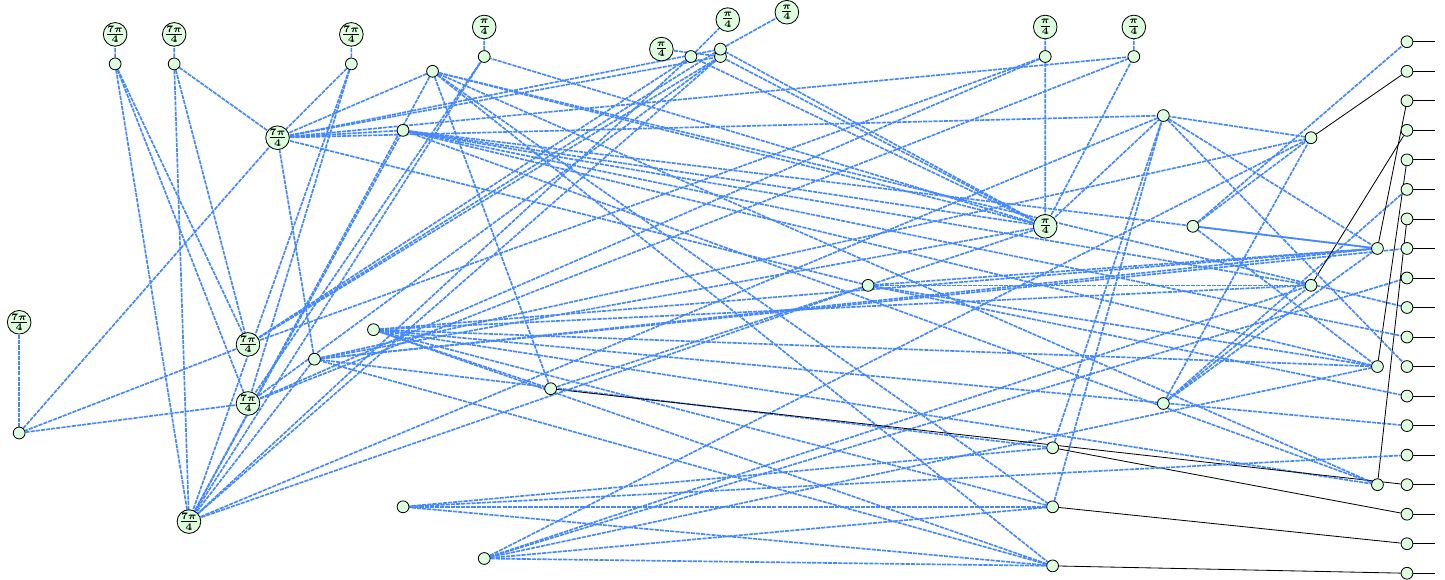}}};
  \node[left=0.3cm of A3] (plus3) {$+$};
\end{tikzpicture}
\end{equation}

\noindent Similar to the $d=3$ example, the terms lying on the top left and bottom right (also bottom left and top right) of \eqref{eq:naive_d5_term} are identical ZX-diagrams. Hence we can reduce the total number of terms to $2$.

We then perform a BSS decomposition\footnote{Using $\mathtt{zx.simulate.find\_stabilizer\_decomp(g)}$ in $\mathtt{pyzx}$ to find the BSS decomposition.} on each of the $2$ surviving terms separately and found that each term can be represented by a sum of $36$ Clifford ZX-diagrams. In total, the $d=5$ variant resultant state has a stabiliser decomposition of $36\times 4 = 72$ terms. 

\subsection{Re-using the $d=3$ stabiliser decomposition then perform procedurally optimised cutting}
\label{subsec:tag}
Can we do better than $72$ terms? We can replace the contents inside the dashed black box in \eqref{eq:MSC_d_5_full} with the final stabiliser decomposition of the $d=3$ variant from \eqref{eq:cut2_after_simp}. Hence, the ZX-diagram from \eqref{eq:MSC_d_5_full} can be written as a sum of $2$ ZX-diagrams:
\begin{equation}
\centering
\label{eq:cutcut0123}
\begin{tikzpicture}[every node/.style={anchor=center},baseline=(A1.center)]
    \node (A1) {\scalebox{0.15}{\includegraphics{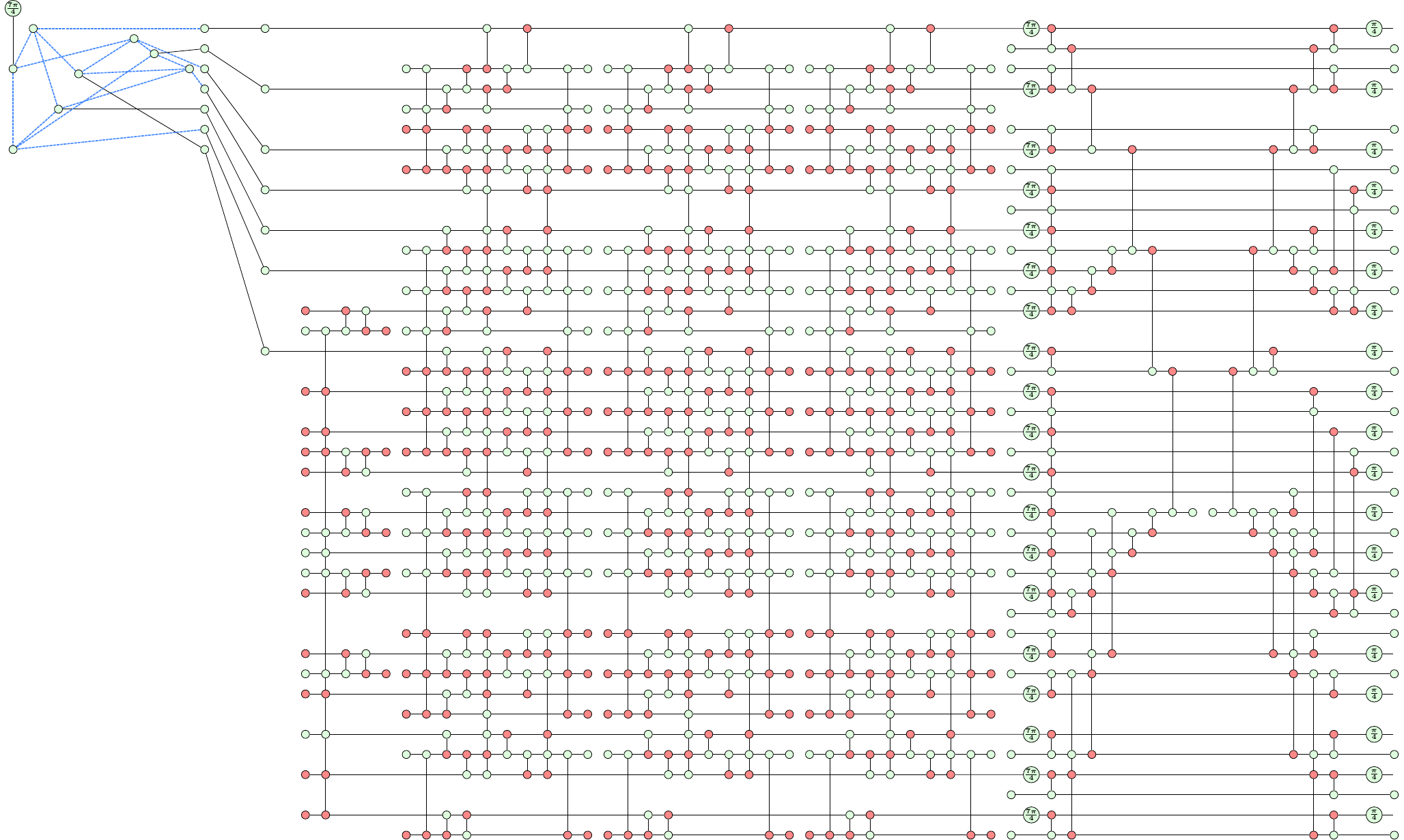}}};
  \node[right=0.3cm of A1] (plus1) {$+$};
    \node[right=0.3cm of plus1] (A2) {\scalebox{0.15}{\includegraphics{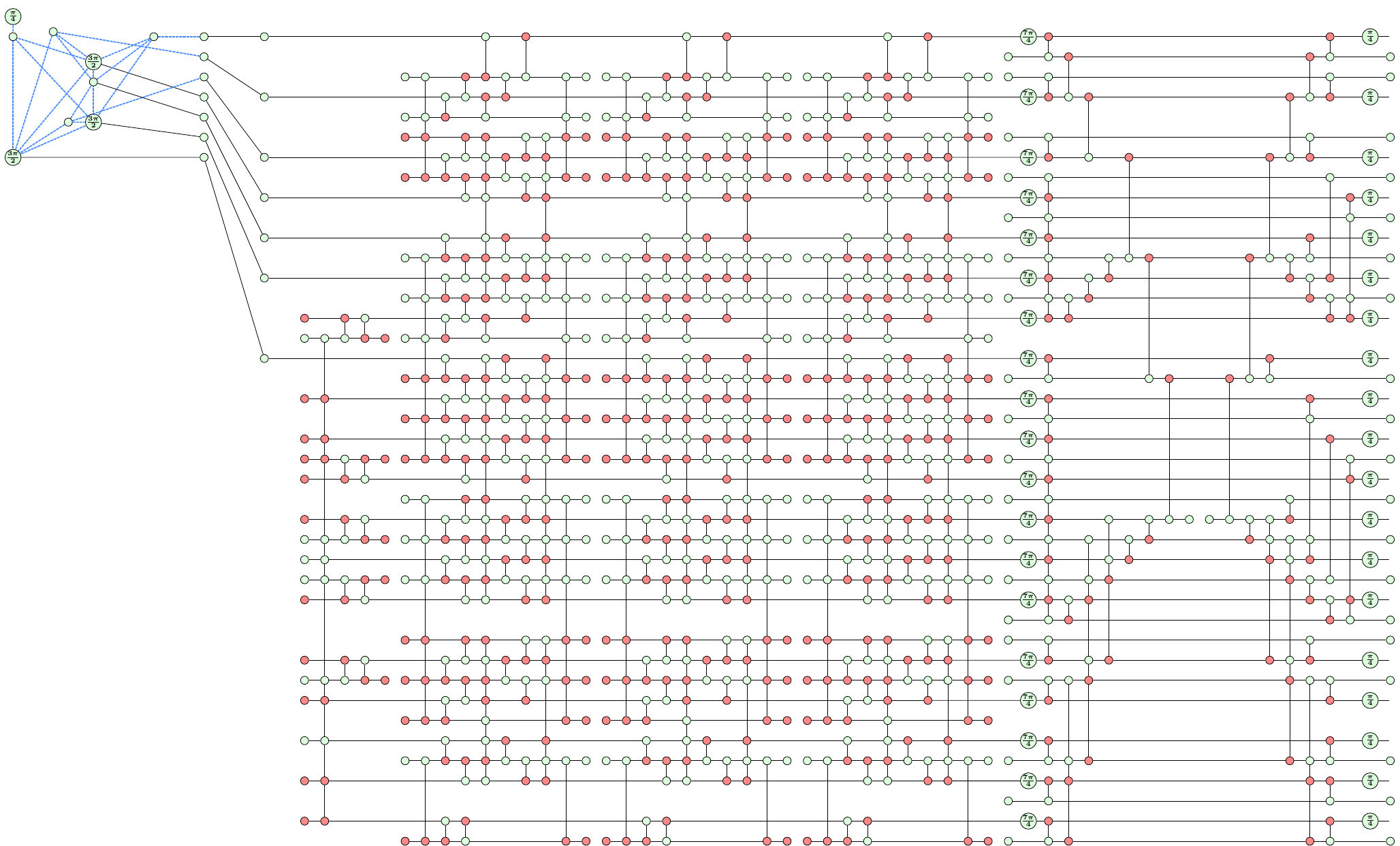}}};
\end{tikzpicture} \ .
\end{equation}

\noindent We can then simplify these $2$ ZX-diagram via re-writes and apply the procedural cutting to this sum of $2$ ZX-diagram. Cutting at $6$ different spiders across the forkings of the $2$ initial terms. This will result in $8$ ZX-diagrams, each with a single $\scalebox{1}{
\begin{tikzpicture}
    \begin{pgfonlayer}{nodelayer}
        \node [style=Z phase dot] (0) at (-2.00, -0.00) {$\frac{\pi}{4}$};
        \node [style=none] (1) at (-1.00, -0.00) {};
    \end{pgfonlayer}
    \begin{pgfonlayer}{edgelayer}
        \draw (0) to (1);
    \end{pgfonlayer}
\end{tikzpicture}
}$/$\scalebox{1}{
\begin{tikzpicture}
    \begin{pgfonlayer}{nodelayer}
        \node [style=Z phase dot] (0) at (-2.00, -0.00) {$\frac{7\pi}{4}$};
        \node [style=none] (1) at (-1.00, -0.00) {};
    \end{pgfonlayer}
    \begin{pgfonlayer}{edgelayer}
        \draw (0) to (1);
    \end{pgfonlayer}
\end{tikzpicture}
}$ spider after further ZX re-writes and simplifications. Of these $8$ ZX-diagrams, we have $4$ pairs of identical ZX-diagrams. Hence further reducing the total number of terms to $4$ ZX-diagrams, each with a single $T$-count. Using the same trick from \eqref{eq:single_magic}, we have a total of $2\times 4 = 8$ total Clifford terms in this stabiliser decomposition. The tikz files to all $4$ terms with a single $\scalebox{1}{
\begin{tikzpicture}
    \begin{pgfonlayer}{nodelayer}
        \node [style=Z phase dot] (0) at (-2.00, -0.00) {$\frac{\pi}{4}$};
        \node [style=none] (1) at (-1.00, -0.00) {};
    \end{pgfonlayer}
    \begin{pgfonlayer}{edgelayer}
        \draw (0) to (1);
    \end{pgfonlayer}
\end{tikzpicture}
}$/$\scalebox{1}{
\begin{tikzpicture}
    \begin{pgfonlayer}{nodelayer}
        \node [style=Z phase dot] (0) at (-2.00, -0.00) {$\frac{7\pi}{4}$};
        \node [style=none] (1) at (-1.00, -0.00) {};
    \end{pgfonlayer}
    \begin{pgfonlayer}{edgelayer}
        \draw (0) to (1);
    \end{pgfonlayer}
\end{tikzpicture}
}$ spider can be found in the arXiv tex source folder - $\mathtt{d5\_additional\_terms}$. 

\section{Summary and discussions}
We have shown that resultant states from the $d=3$ and $d=5$ variants of magic state cultivation circuits \cite{gidney2024magicstatecultivationgrowing}\footnote{Up to the end of the `cultivation' stage.} can be represented as a sum of $4$ and $8$ Clifford ZX-diagrams respectively using the cutting decomposition from \cite{Sutcliffe_2024}. We summarise the different decompositions in the literature applied to the MSC circuits in table \ref{tab:summary}.
\begin{table}[H]
    \centering
\begin{tabular}{ |>{\centering\arraybackslash}p{1.5cm}|>{\centering\arraybackslash}p{1.5cm}||>{\centering\arraybackslash}p{2.45cm}|>{\centering\arraybackslash}p{2.25cm}|>{\centering\arraybackslash}p{2.25cm}|>{\centering\arraybackslash}p{2.25cm}| }
 \hline
 \multicolumn{2}{|c||}{} & \multicolumn{4}{c|}{\textbf{Terms in decomposition}} \\
 \hline
 \textbf{MSC variant} & \textbf{$T$-count ($t$)} & \textbf{BSS (using \cite{kissinger2020Pyzx,BSS2016})} & \textbf{Magic cat state \cite{Codsi2022Masters}\tablefootnote{Terms in the tables from \cite{Codsi2022Masters} are expressed as the number of ZX-diagrams with a single $T/T^{\dagger}$ spider. So our estimates is $2\times$ of that (per magic cat state used) when $\ket{T}^{\otimes (m-1)}$ is expressed as sum of purely Clifford ZX-diagrams.}} & \textbf{Cutting (this work)} & \textbf{Worst case ($2^t$)} \\
 \hline 
 $d=3$ & $15$  & $68$ & $108$ & $4$ & $32{,} 768$ \\
 \hline
 $d=5$ & $53$  & $\approx 29{,}176{,}466$\tablefootnote{The simulation did not finish running after $2$ hours, we resorted to an estimate of $\approx 7^{53/6}$ used in \cite{Sutcliffe_2024}.} & $6{,}377{,}292$\tablefootnote{Combination of $\ket{\text{cat}_{38}}$ and $\ket{\text{cat}_{17}}$.} & $8$  & $9.01 \times 10^{15}$\\
 \hline
\end{tabular}
    \caption{A summary of various stabiliser decompositions applied to the MSC circuits.}
    \label{tab:summary}
\end{table}

\noindent We can see that the cutting decomposition can produce $8\times 10^{5}$ times fewer terms compared to the magic cat state decomposition\footnote{Magic cat state decomposition usually performs better than BSS \cite{Sutcliffe_2024,Sutcliffe2025thesis}.} in the $d=5$ case. Beyond that, $4$ and $8$ terms are still reasonable number of terms to perform computation by hand. 

We wish to emphasise that the resultant state to both the $d=3$ and $d=5$ MSC circuits are just encoded magic states: $\ket{\bar{T}} \propto \ket{\bar{0}} + e^{i\pi/4}\ket{\bar{1}}$. This is a superposition of $2$ Clifford states $\ket{\bar{0}}$ and $\ket{\bar{1}}$, beating the number of terms obtained by the cutting decomposition. Nonetheless, the cutting decomposition presents an interesting avenue for exploration. Can we prune the total number of terms in full syndrome decoding of MSC in a clever way? To replicate the same logical error rates obtained from full state-vector simulation?

The authors from \cite{gidney2024magicstatecultivationgrowing} simulated most of the larger circuits by replacing the $T$ gates entirely with $S$ gates, but noticed a slight performance difference when benchmarked against brute-force state vector simulations for smaller circuit. We hope that this study can be a baby starting step towards simulating these circuits exactly and also highlight the versatility of ZX-calculus.

\section{Acknowledgements}
The authors acknowledge discussions with Austin Fowler and Aleks Kissinger regarding non-Clifford ZX-diagram Pauli webs in an earlier iteration of this work. Kwok Ho Wan wishes to thank his wife for all the computer hardware and coding related support. Zhenghao Zhong is supported by the ERC Consolidator Grant \# 864828 ``Algebraic Foundations of Supersymmetric Quantum Field Theory'' (SCFTAlg). The authors would like to acknowledge the use of the University of Oxford Advanced Research Computing (ARC) facility in carrying out previous failed iterations/explorations of this work. We also wish to thank Matthew Sutcliffe for spotting a typo in the full $d=5$ circuit BSS decomposition estimation (table \ref{tab:summary}\footnote{See scirate comment: \url{https://scirate.com/arxiv/2509.01224\#2628}.}).

Tikz diagram in \eqref{eq:bss_decomp} is taken from from \cite{Sutcliffe_2024} while \eqref{eq:2-T-decomposition}, \eqref{eq:cat-n}, \eqref{eq:cat-6-decomp}, \eqref{eq:magic-from-cat}, \eqref{eq:spider_decomp_abs},  \eqref{eq:picom_grouped} and \eqref{eq:single_magic} are taken from \cite{Kiss2022}. Packages such as $\mathtt{zxlive}$, $\mathtt{pyzx}$ \cite{kissinger2020Pyzx}, magic state cultivation simulation python code \cite{gidneyy2024cultivationdata} and cutting stabiliser decomposition python code \cite{Sutcliffe_2024} have been extensively used in the preparation of this manuscript.
 
\bibliography{main}

\begin{thebibliography}{16}
\providecommand{\natexlab}[1]{#1}
\providecommand{\url}[1]{\texttt{#1}}
\expandafter\ifx\csname urlstyle\endcsname\relax
  \providecommand{\doi}[1]{doi: #1}\else
  \providecommand{\doi}{doi: \begingroup \urlstyle{rm}\Url}\fi

\bibitem[Gottesman(1998)]{gottesman1998heisenbergrepresentationquantumcomputers}
Daniel Gottesman.
\newblock The {H}eisenberg {R}epresentation of {Q}uantum {C}omputers, 1998.
\newblock URL \url{https://arxiv.org/abs/quant-ph/9807006}.

\bibitem[Aaronson and Gottesman(2004)]{Aaronson_2004}
Scott Aaronson and Daniel Gottesman.
\newblock Improved simulation of stabilizer circuits.
\newblock \emph{Physical Review A}, 70\penalty0 (5), November 2004.
\newblock ISSN 1094-1622.
\newblock \doi{10.1103/physreva.70.052328}.
\newblock URL \url{http://dx.doi.org/10.1103/PhysRevA.70.052328}.

\bibitem[Bravyi and Gosset(2016)]{PhysRevLett.116.250501}
Sergey Bravyi and David Gosset.
\newblock Improved {C}lassical {S}imulation of {Q}uantum {C}ircuits {D}ominated by {C}lifford {G}ates.
\newblock \emph{Phys. Rev. Lett.}, 116:\penalty0 250501, Jun 2016.
\newblock \doi{10.1103/PhysRevLett.116.250501}.
\newblock URL \url{https://link.aps.org/doi/10.1103/PhysRevLett.116.250501}.

\bibitem[Bremner et~al.(2010)Bremner, Jozsa, and Shepherd]{Bremner_2010}
Michael~J. Bremner, Richard Jozsa, and Dan~J. Shepherd.
\newblock Classical simulation of commuting quantum computations implies collapse of the polynomial hierarchy.
\newblock \emph{Proceedings of the Royal Society A: Mathematical, Physical and Engineering Sciences}, 467\penalty0 (2126):\penalty0 459–472, August 2010.
\newblock ISSN 1471-2946.
\newblock \doi{10.1098/rspa.2010.0301}.
\newblock URL \url{http://dx.doi.org/10.1098/rspa.2010.0301}.

\bibitem[Sutcliffe and Kissinger(2024)]{Sutcliffe_2024}
Matthew Sutcliffe and Aleks Kissinger.
\newblock Procedurally {O}ptimised {ZX}-{D}iagram {C}utting for {E}fficient {T}-{D}ecomposition in {C}lassical {S}imulation.
\newblock \emph{Electronic Proceedings in Theoretical Computer Science}, 406:\penalty0 63–78, August 2024.
\newblock ISSN 2075-2180.
\newblock \doi{10.4204/eptcs.406.3}.
\newblock URL \url{http://dx.doi.org/10.4204/EPTCS.406.3}.

\bibitem[Gidney et~al.(2024)Gidney, Shutty, and Jones]{gidney2024magicstatecultivationgrowing}
Craig Gidney, Noah Shutty, and Cody Jones.
\newblock Magic state cultivation: growing {T} states as cheap as {CNOT} gates, 2024.
\newblock URL \url{https://arxiv.org/abs/2409.17595}.

\bibitem[Bravyi and Kitaev(2005)]{Bravyi_2005}
Sergey Bravyi and Alexei Kitaev.
\newblock Universal quantum computation with ideal {C}lifford gates and noisy ancillas.
\newblock \emph{Physical Review A}, 71\penalty0 (2), February 2005.
\newblock ISSN 1094-1622.
\newblock \doi{10.1103/physreva.71.022316}.
\newblock URL \url{http://dx.doi.org/10.1103/PhysRevA.71.022316}.

\bibitem[Litinski(2019)]{Litinski_2019}
Daniel Litinski.
\newblock Magic {S}tate {D}istillation: {N}ot as {C}ostly as {Y}ou {T}hink.
\newblock \emph{Quantum}, 3:\penalty0 205, December 2019.
\newblock ISSN 2521-327X.
\newblock \doi{10.22331/q-2019-12-02-205}.
\newblock URL \url{http://dx.doi.org/10.22331/q-2019-12-02-205}.

\bibitem[Kissinger and van~de Wetering(2020)]{kissinger2020Pyzx}
Aleks Kissinger and John van~de Wetering.
\newblock {PyZX: Large Scale Automated Diagrammatic Reasoning}.
\newblock In Bob Coecke and Matthew Leifer, editors, \emph{{\rm Proceedings 16th International Conference on} Quantum Physics and Logic, {\rm Chapman University, Orange, CA, USA., 10-14 June 2019}}, volume 318 of \emph{Electronic Proceedings in Theoretical Computer Science}, pages 229--241. Open Publishing Association, 2020.
\newblock \doi{10.4204/EPTCS.318.14}.

\bibitem[Kissinger and van~de Wetering()]{KissingerWetering2024Book}
Aleks Kissinger and John van~de Wetering.
\newblock \emph{{P}icturing {Q}uantum {S}oftware: {A}n {I}ntroduction to the {ZX}-{C}alculus and {Q}uantum {C}ompilation}.

\bibitem[Kissinger et~al.(2022)Kissinger, van~de Wetering, and Vilmart]{Kiss2022}
Aleks Kissinger, John van~de Wetering, and Renaud Vilmart.
\newblock Classical simulation of quantum circuits with partial and graphical stabiliser decompositions.
\newblock Schloss Dagstuhl – Leibniz-Zentrum für Informatik, 2022.
\newblock \doi{10.4230/LIPICS.TQC.2022.5}.
\newblock URL \url{https://drops.dagstuhl.de/entities/document/10.4230/LIPIcs.TQC.2022.5}.

\bibitem[Qassim et~al.(2021)Qassim, Pashayan, and Gosset]{Qassim_2021}
Hammam Qassim, Hakop Pashayan, and David Gosset.
\newblock Improved upper bounds on the stabilizer rank of magic states.
\newblock \emph{Quantum}, 5:\penalty0 606, December 2021.
\newblock ISSN 2521-327X.
\newblock \doi{10.22331/q-2021-12-20-606}.
\newblock URL \url{http://dx.doi.org/10.22331/q-2021-12-20-606}.

\bibitem[Bravyi et~al.(2016)Bravyi, Smith, and Smolin]{BSS2016}
Sergey Bravyi, Graeme Smith, and John~A. Smolin.
\newblock Trading {C}lassical and {Q}uantum {C}omputational {R}esources.
\newblock \emph{Phys. Rev. X}, 6:\penalty0 021043, Jun 2016.
\newblock \doi{10.1103/PhysRevX.6.021043}.
\newblock URL \url{https://link.aps.org/doi/10.1103/PhysRevX.6.021043}.

\bibitem[Codsi(2022)]{Codsi2022Masters}
Julien Codsi.
\newblock {Cutting-Edge Graphical Stabiliser Decompositions for Classical Simulation of Quantum Circuits}.
\newblock Master's thesis, University of Oxford, 2022.

\bibitem[Sutcliffe(2025)]{Sutcliffe2025thesis}
Matthew Sutcliffe.
\newblock \emph{Novel {M}ethods for {C}lassical {S}imulation of {Q}uantum {C}ircuits via {ZX}-Calculus}.
\newblock PhD thesis, University of Oxford, 2025.

\bibitem[Gidney(2024)]{gidneyy2024cultivationdata}
Craig Gidney.
\newblock {Data for "Magic state cultivation: growing T states as cheap as CNOT gates"}.
\newblock \emph{Zenodo}, September 2024.
\newblock \doi{10.5281/zenodo.13777072}.

\end{thebibliography}

\appendix 

\section{Equality for $d=3$ MSC}

\begin{equation}
\centering
\begin{tikzpicture}[baseline=(m.center)]
  \matrix (m) [matrix of nodes,
               row sep=3mm, column sep=2mm,
               nodes={anchor=center, inner sep=0pt}]
  {
    \scalebox{0.25}{\includegraphics{MSC_d_3_full.pdf}} & = &
    \scalebox{0.6}{\begin{tikzpicture}
    \begin{pgfonlayer}{nodelayer}
        \node [style=none] (1) at (24.00, -4.00) {};
        \node [style=Z dot] (2) at (18.00, -0.50) {};
        \node [style=Z phase dot] (3) at (12.00, 1.00) {$\frac{7\pi}{4}$};
        \node [style=none] (4) at (24.00, -6.00) {};
        \node [style=none] (5) at (24.00, -1.00) {};
        \node [style=none] (6) at (24.00, -0.00) {};
        \node [style=Z dot] (7) at (23.00, -3.00) {};
        \node [style=none] (8) at (24.00, -5.00) {};
        \node [style=Z dot] (9) at (23.00, -2.00) {};
        \node [style=Z dot] (10) at (12.00, -2.00) {};
        \node [style=Z dot] (11) at (19.00, -1.25) {};
        \node [style=Z dot] (12) at (23.00, -6.00) {};
        \node [style=Z dot] (13) at (13.00, -0.00) {};
        \node [style=Z dot] (14) at (23.00, -1.00) {};
        \node [style=none] (15) at (24.00, -2.00) {};
        \node [style=Z dot] (16) at (23.00, -4.00) {};
        \node [style=Z dot] (17) at (20.75, -2.00) {};
        \node [style=Z dot] (18) at (12.00, -6.00) {};
        \node [style=none] (19) at (24.00, -3.00) {};
        \node [style=Z dot] (20) at (14.25, -4.00) {};
        \node [style=Z dot] (21) at (23.00, -5.00) {};
        \node [style=Z dot] (22) at (23.00, -0.00) {};
        \node [style=Z dot] (23) at (15.25, -2.25) {};
    \end{pgfonlayer}
    \begin{pgfonlayer}{edgelayer}
        \draw (1) to (16);
        \draw [style=hadamard edge] (2) to (23);
        \draw [style=hadamard edge] (2) to (11);
        \draw [style=hadamard edge] (2) to (9);
        \draw [style=hadamard edge] (2) to (10);
        \draw (3) to (10);
        \draw (4) to (12);
        \draw (5) to (14);
        \draw (6) to (22);
        \draw (7) to (19);
        \draw [style=hadamard edge] (7) to (17);
        \draw (8) to (21);
        \draw (9) to (15);
        \draw [style=hadamard edge] (10) to (13);
        \draw [style=hadamard edge] (10) to (18);
        \draw (11) to (14);
        \draw [style=hadamard edge] (11) to (17);
        \draw [style=hadamard edge] (11) to (18);
        \draw (12) to (23);
        \draw [style=hadamard edge] (13) to (20);
        \draw [style=hadamard edge] (13) to (23);
        \draw [style=hadamard edge] (13) to (22);
        \draw (16) to (20);
        \draw [style=hadamard edge] (17) to (20);
        \draw [style=hadamard edge] (17) to (23);
        \draw [style=hadamard edge] (18) to (20);
        \draw [style=hadamard edge] (18) to (21);
    \end{pgfonlayer}
\end{tikzpicture}} & + &
    \scalebox{0.6}{\begin{tikzpicture}
    \begin{pgfonlayer}{nodelayer}
        \node [style=Z dot] (0) at (27.50, -0.00) {};
        \node [style=Z dot] (1) at (31.50, -2.25) {};
        \node [style=Z dot] (2) at (29.50, 0.25) {};
        \node [style=Z dot] (3) at (34.50, -0.00) {};
        \node [style=Z dot] (4) at (30.25, -4.25) {};
        \node [style=Z phase dot] (5) at (27.50, -6.00) {$\frac{3\pi}{2}$};
        \node [style=Z phase dot] (6) at (31.50, -4.25) {$\frac{3\pi}{2}$};
        \node [style=Z phase dot] (7) at (31.50, -1.25) {$\frac{3\pi}{2}$};
        \node [style=Z phase dot] (8) at (27.50, 1.00) {$\frac{\pi}{4}$};
        \node [style=Z dot] (9) at (37.00, 0.00) {};
        \node [style=Z dot] (10) at (37.00, -1.00) {};
        \node [style=Z dot] (11) at (37.00, -2.00) {};
        \node [style=Z dot] (12) at (37.00, -3.00) {};
        \node [style=Z dot] (13) at (37.00, -4.00) {};
        \node [style=Z dot] (14) at (37.00, -5.00) {};
        \node [style=Z dot] (15) at (37.00, -6.00) {};
        \node [style=none] (16) at (38.00, 0.00) {};
        \node [style=none] (17) at (38.00, -1.00) {};
        \node [style=none] (18) at (38.00, -2.00) {};
        \node [style=none] (19) at (38.00, -3.00) {};
        \node [style=none] (20) at (38.00, -4.00) {};
        \node [style=none] (21) at (38.00, -5.00) {};
        \node [style=none] (22) at (38.00, -6.00) {};
    \end{pgfonlayer}
    \begin{pgfonlayer}{edgelayer}
        \draw [style=hadamard edge] (0) to (8);
        \draw [style=hadamard edge] (0) to (6);
        \draw [style=hadamard edge] (0) to (7);
        \draw [style=hadamard edge] (0) to (5);
        \draw [style=hadamard edge] (1) to (4);
        \draw [style=hadamard edge] (1) to (2);
        \draw [style=hadamard edge] (1) to (3);
        \draw (1) to (13);
        \draw [style=hadamard edge] (2) to (7);
        \draw [style=hadamard edge] (2) to (5);
        \draw [style=hadamard edge] (2) to (10);
        \draw [style=hadamard edge] (3) to (6);
        \draw [style=hadamard edge] (3) to (7);
        \draw [style=hadamard edge] (3) to (9);
        \draw [style=hadamard edge] (4) to (5);
        \draw [style=hadamard edge] (4) to (6);
        \draw [style=hadamard edge] (4) to (11);
        \draw [style=hadamard edge] (5) to (7);
        \draw [style=hadamard edge] (5) to (6);
        \draw (5) to (15);
        \draw [style=hadamard edge] (6) to (7);
        \draw (6) to (14);
        \draw (7) to (12);
        \draw (9) to (16);
        \draw (10) to (17);
        \draw (11) to (18);
        \draw (12) to (19);
        \draw (13) to (20);
        \draw (14) to (21);
        \draw (15) to (22);
    \end{pgfonlayer}
\end{tikzpicture}} \\
  };
\end{tikzpicture}
\label{eq:large_d3}
\end{equation}

\section{Equality for $d=5$ MSC}

\begin{equation}
\centering
\begin{tikzpicture}[baseline=(m.center)]
  \matrix (m) [matrix of nodes,
               row sep=3mm, column sep=2mm,
               nodes={anchor=center, inner sep=0pt}]
  {
    \scalebox{0.1}{\includegraphics{d5_dashed.pdf}} & = &
    \scalebox{0.1}{\includegraphics{cutcut0.pdf}} & + &
    \scalebox{0.1}{\includegraphics{cutcut1.pdf}} \\
    ~ & = &
    \scalebox{0.2}{\includegraphics{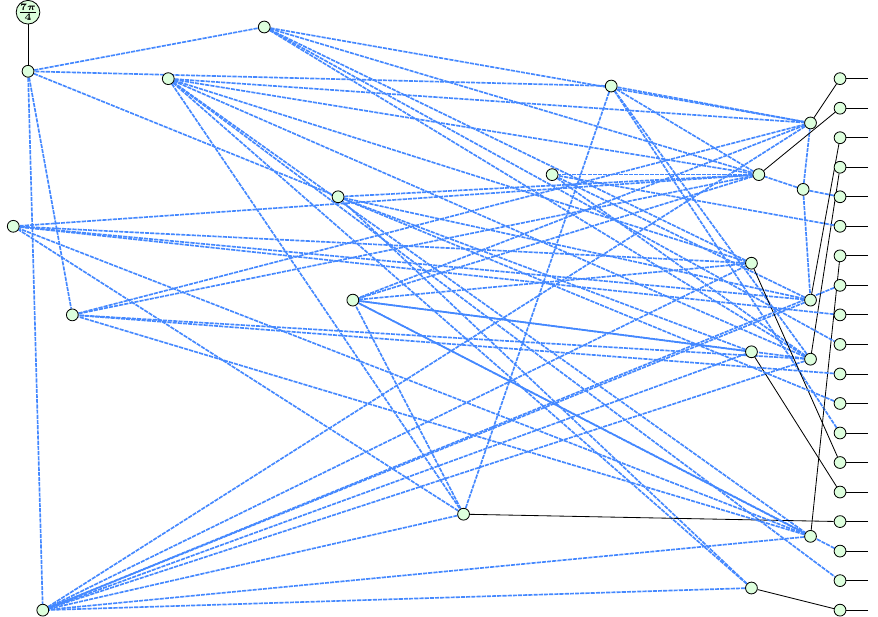}} & + &
    \scalebox{0.2}{\includegraphics{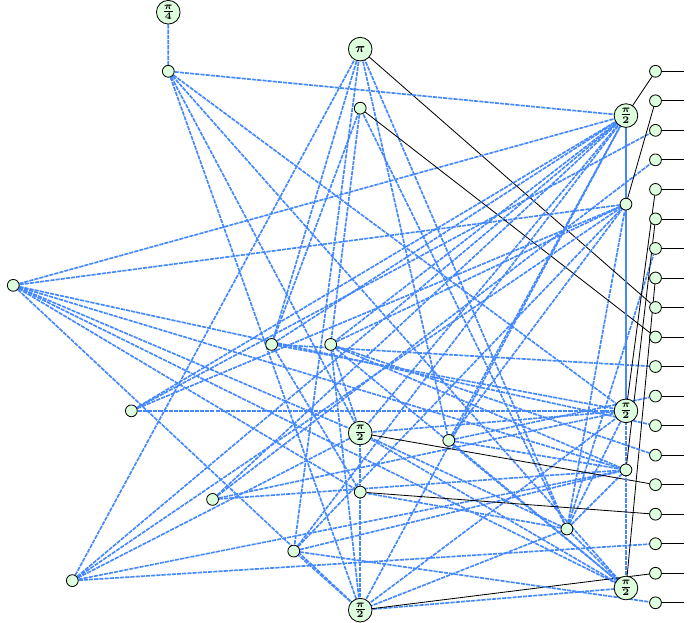}} \\
    ~ & + &
    \scalebox{0.2}{\includegraphics{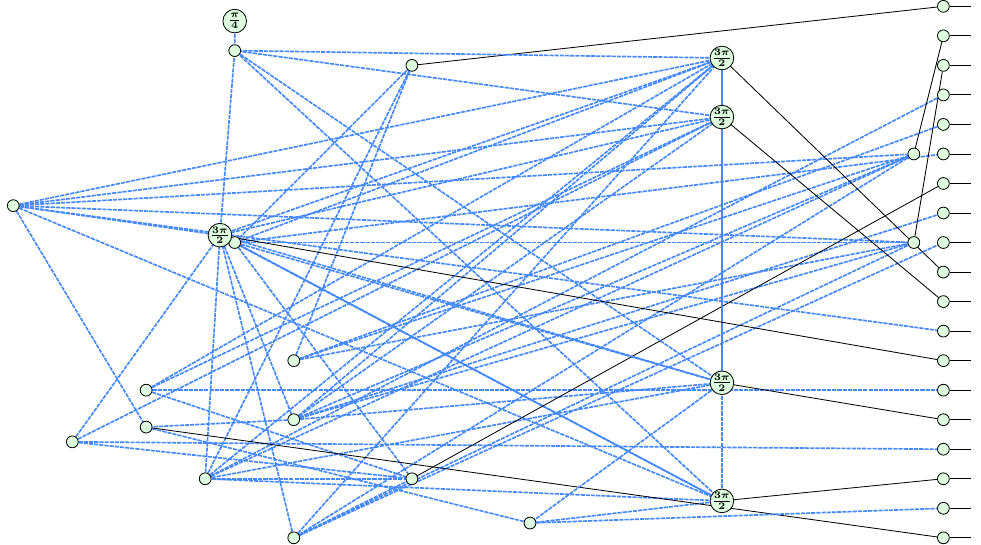}} & + &
    \scalebox{0.2}{\includegraphics{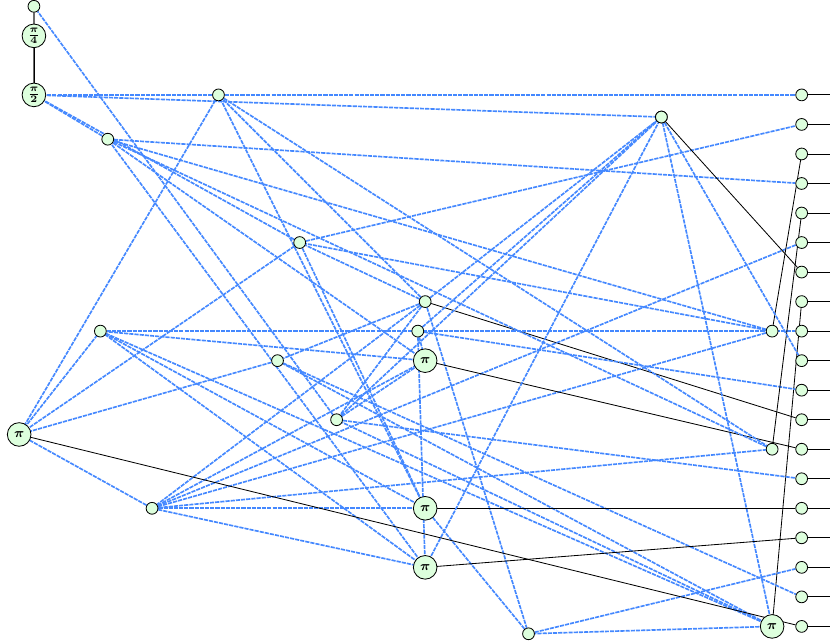}} \\
  };
\end{tikzpicture}
\label{eq:large_d5}
\end{equation}

\end{document}